\documentclass[preprint]{emulateapj}%aastex}%{
\usepackage{graphicx}
\usepackage{amsmath}
\usepackage[mathscr]{eucal}
\usepackage{gensymb}
\usepackage{subfigure}
\usepackage{booktabs}
\usepackage{multirow}
\usepackage{color}

\begin{document}
%
% 		** TITLE **
%============================
\title 
{On-sky demonstration of low-order wavefront sensing and control with focal plane phase mask coronagraphs}
\author
{
Garima Singh$^{1,2}$, Julien Lozi$^{1}$, Olivier Guyon$^{1}$, Pierre Baudoz $^{2}$, Nemanja Jovanovic$^{1}$, Frantz Martinache$^{3}$, Tomoyuki Kudo$^{1}$, Eugene Serabyn$^{4}$, Jonas Kuhn$^{5}$\\
$^1$\small National Astronomical Observatory of Japan, Subaru Telescope, 650 N A'Ohoku Place, Hilo, HI  96720, USA\\
$^2$\small Lesia, Observatoire de Paris-Meudon, 5 Place Jules Janssen, F-92195 Meudon Cedex, France\\
$^3$\small Observatoire de la C\^ote d'Azur, Boulevard de l'Observatoire, 06300 Nice, France\\
$^4$\small Jet Propulsion Laboratory, California Institute of Technology, 4800 Oak Grove Drive, Pasadena, CA 91109-8099, USA\\
$^5$\small Institute for Astronomy, Swiss Federal Institute of Technology, Wolfgang-Pauli-Strasse 27, CH-8093 Zurich, Switzerland\\
}
\email{singh@naoj.org} 
%
%			**ABSTRACT**
%================================

\begin{abstract}

The ability to characterize exoplanets by spectroscopy of their atmospheres requires direct imaging techniques to isolate planet signal from the bright stellar glare. One of the limitations with the direct detection of exoplanets, either with ground- or space-based coronagraphs, is pointing errors and other low-order wavefront aberrations. The coronagraphic detection sensitivity at the diffraction limit therefore depends on how well low-order aberrations upstream of the focal plane mask are corrected. To prevent starlight leakage at the inner working angle of a phase mask coronagraph, we have introduced a Lyot-based low-order wavefront sensor (LLOWFS), which senses aberrations using the rejected starlight diffracted at the Lyot plane. In this paper, we present the implementation, testing and results of LLOWFS on the Subaru Coronagraphic Extreme Adaptive Optics system (SCExAO) at the Subaru Telescope. 

We have controlled thirty-five Zernike modes of a H-band vector vortex coronagraph in the laboratory and ten Zernike modes on sky with an integrator control law. We demonstrated a closed-loop pointing residual of 0.02~mas in the laboratory and 0.15~mas on sky for data sampled using the minimal 2-second exposure time of the science camera. We have also integrated the LLOWFS in the visible high-order control loop of SCExAO, which in closed-loop operation has validated the correction of the non-common path pointing errors between the infrared science channel and the visible wavefront sensing channel with pointing residual of 0.23~mas on sky. 

\end{abstract}

%
%		**1. INTRODUCTION**
%================================

\section{Introduction}

One of the goals of the next generation of ground- and space-based missions is the direct detection and spectrophotometric characterization of rocky-type exoplanets in the habitable zone (HZ) of a parent star. The scientific motivation is to study the chemical compositions of their atmospheres to search for biosignatures. Disentangling rocky-type extrasolar planets from M-type and solar-type star at 10~parsec requires the angular resolution and sensitivity of a 30-m telescope from the ground and 2-4 meters telescope in space respectively. However, resolution alone is not sufficient enough for their detection in the HZ. The direct imaging of such exoplanets is challenged by the ability of identifying planet signal above the bright stellar background at small angular separation, which therefore requires high contrast imaging (HCI) near the diffraction limit.

Coronagraphs are used to block the starlight and suppress the diffraction effects of the telescope, making the planet signal more accessible. Small inner working angle (IWA) coronagraphs can reach to within the first couple of Airy rings of the star. However, the exploitation of this region relies on the ability of efficiently controlling and calibrating the residual low-order wavefront errors \citep{wferror}. These aberrations occurring upstream of a focal plane mask (FPM) are a common issue for both ground- and space-based coronagraphs, which result in starlight leaking around the coronagraphic mask. The aim of this paper is to present the results of a unique low-order wavefront sensor applicable to phase mask coronagraphs (PMCs), including the vortex coronagraph, with which it is tested here.

First efforts have been made to reduce the quasi-static pointing aberrations at Palomar well-corrected subaperture (WCS), on the Hale telescope, and achieved a residual of 0.02 $\lambda/D$ (6~mas) with a vortex coronagraph \citep{eugene}. The current ground-based extreme adaptive optics (ExAO) instruments such as Gemini Planet Imager \citep[GPI,][]{gpi} at the Gemini Observatory and Spectro-Polarimetric High-contrast Exoplanet Research \citep[SPHERE,][]{sphere} at the Very Large Telescope are now predictively correcting the dynamic low-order wavefront aberrations.

GPI is equipped with a 7$\times$7 low-order Shack-Hartmann (SH) wavefront sensor that has demonstrated the corrections of the non-common path aberrations down to $<$~5~nm root mean square (RMS) for spatial frequencies $<$~3 cycles/pupil under simulated turbulence. By implementing a Linear Quadratic Gaussian algorithm \cite[LQG,][]{lqg} in the AO system, they have demonstrated on-sky corrections of common-path vibrations at 60, 120 and 180 Hz to under 1 mas per axis for tip-tilt residuals and a reduction of focus aberration down to 3~nm~RMS wavefront error at the 60~Hz peak \citep{Poyneer}.  

SPHERE's SAXO (SPHERE AO for eXoplanet Observation) uses a 40$\times$40 visible SH wavefront sensor and demonstrated an on-sky residual jitter of 11~mas with an integrator controller and 9~mas with an LQG algorithm \citep{Petit}.

The Subaru Coronagraphic ExAO \cite[SCExAO,][]{scexao} instrument at the Subaru Telescope, the Exoplanetary Circumstellar Environments and Disk Explorer \cite[EXCEDE,][]{belikov, JL} testbed at NASA Ames and the High-Contrast Imaging Testbed \citep[HCIT,][]{kern} at JPL have implemented a coronagraphic low-order wavefront sensor \citep[CLOWFS,][]{clowfs1}, which senses the rejected starlight reflected by the FPM. With the use of a Phase-Induced Amplitude Apodization \citep[PIAA,][]{piaa} coronagraph, residuals $\leq10^{-3}~\lambda/D$ for the tip and tilt modes have been demonstrated in closed-loop in the laboratory operation.

However, these existing solutions are not compatible with the non-reflective PMCs, which are the type of coronagraphs that diffracts the rejected starlight in the post-coronagraphic pupil plane. To address this issue, \cite{Singh1} have introduced the concept of a Lyot-based low-order wavefront sensor (LLOWFS), which senses aberrations using the residual starlight reflected by the Lyot stop. Its first implementation has demonstrated an open loop measurement pointing accuracy of $10^{-2}~\lambda/D$ at 638 nm with a Four Quadrant Phase Mask \citep[FQPM,][]{Rouan} coronagraph. The preliminary implementation of the LLOWFS on the SCExAO instrument has also demonstrated an on-sky closed-loop pointing accuracy of $\sim~7~\times~10^{-3}~\lambda/D$ \citep{Singh2} with a vector vortex coronagraph \citep[VVC,][]{mawet}.

The aim of this paper is to present the laboratory and on-sky results of an improved version of the LLOWFS on the SCExAO instrument. In section~\ref{s:concept}, we remind the reader about the principle of the LLOWFS concept and its integration in the SCExAO instrument. Then, section~\ref{s:llowfsDM} presents the results in laboratory and on-sky for the configuration where the aberrations sensed by the LLOWFS are directly corrected by the Deformable Mirror (DM). Finally, section~\ref{s:pyr} presents the on-sky results for a second configuration where the LLOWFS is integrated in the ExAO loop to correct for the non-common path and chromatic errors between the visible wavefront sensor of the ExAO and the imaging wavelengths. 

 \begin{figure*}
   \centerline{
        \resizebox{0.9\textwidth}{!}{\includegraphics{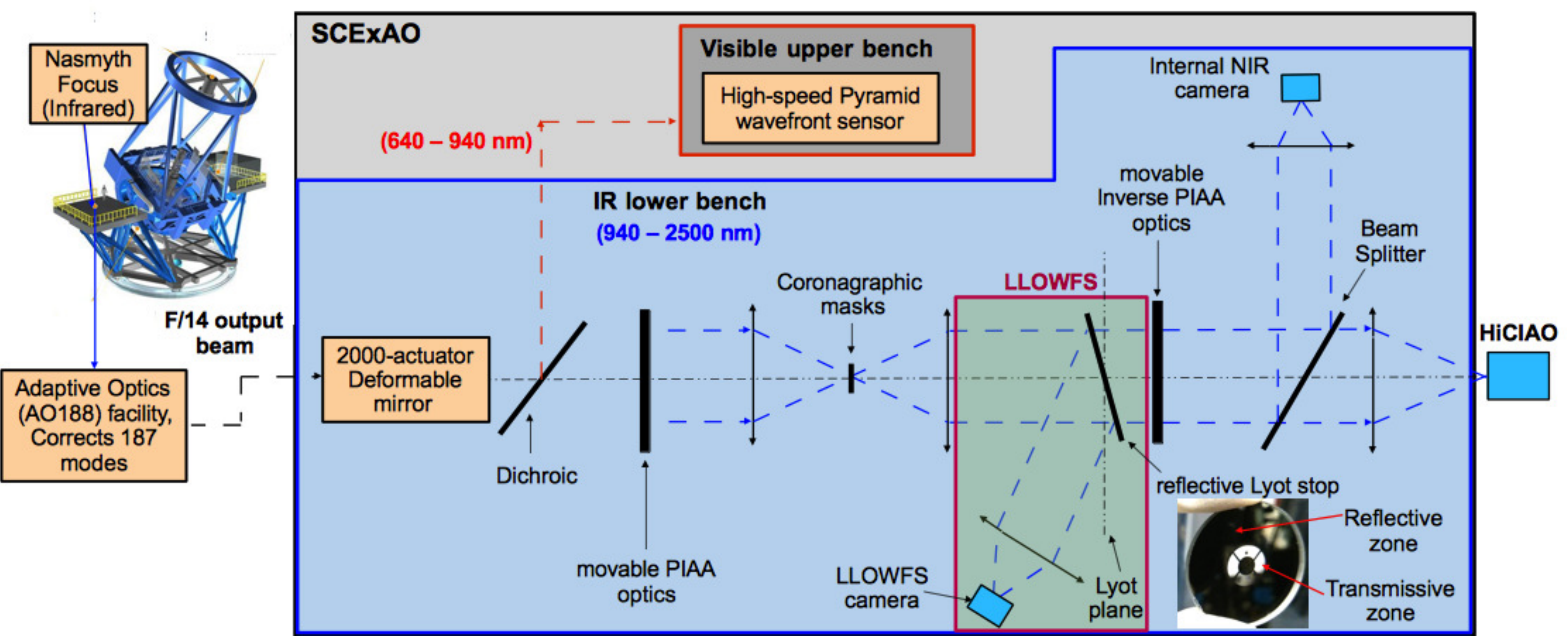}}}
   \caption{Simplified optical ray path of SCExAO. The instrument is situated at the Nasmyth platform of the Subaru Telescope and feeds on the beam from AO188. The output of the instrument goes to the high contrast imager, HiCIAO. SCExAO has two benches: visible and IR. The coronagraphic masks at the focal plane are interchangeable PMCs such as VVC, FQPM and 8OPM. LLOWFS is shown on the IR channel simply requiring a reflective Lyot stop (RLS), relay optics and a detector. The RLS presented in the figure is the Lyot stop designed for the VVC.}
  \label{f:scexao}
\end{figure*}
%

%          **2.) LLOWFS**
%================================

\section{Lyot-based low-order wavefront sensor}
\label{s:concept}

\subsection{Principle}

LLOWFS is a coronagraphic wavefront sensor which is designed to sense the pointing errors and other low-order wavefront aberrations at the IWA of the PMCs. The coronagraphic mask at the focal plane diffracts starlight outside the geometrical pupil in the downstream pupil plane. Unlike conventional coronagraphs, the diffracted starlight in the re-imaged pupil plane, instead of being simply blocked by an opaque Lyot stop, is reflected via a reflective Lyot stop (RLS) towards a re-imaged focal plane. This reflected light is collected by a detector and used to measure the low-order aberrations. 

LLOWFS is a linear wavefront reconstructor that relies on the assumption that if the post-AO wavefront residuals are $\ll$~1~radian~RMS then the intensity variations in the reflected light are a linear combination of the low-order aberrations occurring upstream of the focal plane phase mask.  

An image $I_{R}$ affected by the low-order modes $i$ of amplitude $\boldsymbol{\alpha} = \left(\alpha_1, \alpha_2 \cdots \alpha_n\right)$ is subtracted from a reference image $I_{0}$ and decomposed into a linear combination on a base of orthonormal images $S_{i}$ corresponding to the response of the sensor to the low-order modes. So the difference between an image at any instant and the reference follows the equation

\begin{equation}
I_R (\boldsymbol{\alpha}) - I_0 =  \sum_{i=1}^n \alpha_i S_i.
\end{equation}

The measurements are then used to compute the control commands via an integrator control law. 

This paper focuses on the empirical approach of the LLOWFS only. For a detailed theoretical description, the reader may refer to the publication \cite{Singh1}.
%
%
%          **2.1 SCExAO instrument**
%================================

\subsection{SCExAO instrument with integrated LLOWFS}
\label{s:scexao}

%SCExAO is a versatile high contrast imaging instrument which features an ExAO control loop using a Pyramid wavefront sensor \citep[PyWFS,][]{chris} to measure the high-order wavefront aberrations, a speckle nulling routine to suppress the quasi-static speckles and a LLOWFS to measure the coronagraphic leaks. These different wavefront sensors are implemented on SCExAO to address the issues that degrade the point spread function (PSF) quality. The PyWFS provides a high and stable Strehl ratio, the LLOWFS stabilizes the starlight behind the coronagraphic mask and speckle nulling to reduce the quasi-static aberrations on one half of the field of view. 
%

SCExAO is a versatile high contrast imaging instrument which features an ExAO control loop using a Pyramid wavefront sensor \citep[PyWFS,][]{chris} that provides a high and stable Strehl ratio, a speckle nulling routine to improve the contrast on one half of the field of view and a LLOWFS to stabilize the starlight behind the coronagraphic mask. These different wavefront sensors are implemented on SCExAO to address the issues that degrade the point spread function (PSF) quality: the PyWFS measures the dynamical high-order wavefront aberrations, speckle nulling suppresses the quasi-static speckles and the LLOWFS measures the coronagraphic leaks. This publication focuses only on the LLOWFS and its integration with the PyWFS. More details about the PyWFS and the speckle nulling loop can be found in \cite{scexao}.

The SCExAO instrument is located at the Nasmyth platform of the Subaru Telescope. The instrument is sandwiched between Subaru's 188-actuator adaptive optics facility \citep[AO188,][]{ao} and HiCIAO \citep{Hodapp}, a high-contrast coronographic imager for AO offering angular/spectral/polarization differential imaging modes. Figure \ref{f:scexao} shows the simplified version of the optical ray path on SCExAO which is described as follows. AO188, using the light below 640~nm and correcting 187 modes, stabilizes the PSF with a typical Strehl ratio of 30\% in H band. The AO corrected diffraction-limited F/14 beam is then fed to SCExAO as an input. The beam, collimated by an off-axis parabola (OAP), strikes SCExAO's 2000-actuator DM at the pupil plane. The beam reflected from the DM meets the dichroic that separates the visible light (640 ~-~940~nm) from the Infrared (IR) light (940~-~2500~nm). The visible light is reflected towards the upper bench via a periscope while the IR light is transmitted to the lower bench. The visible upper bench includes a non-modulated PyWFS which is capable of measuring $\sim$~1600 aberrated modes with a frame rate of up to 3.6~kHz at $\sim$~850~nm. The lower IR bench supports the LLOWFS and the speckle nulling control loop working at 1.6 \micron. The bench includes a variety of coronagraphs optimized for very small IWA (1~-~3~$\lambda/D$, i.e. 40-120~mas at 1.6 \micron): PIAA, Shaped pupil \citep{shaped}, VVC, FQPM and eight octant phase mask \citep[8OPM,][]{murakami}. The VVC on SCExAO is a rotating half-waveplate structure that has a vectorial phase spiral. There is a 25-$\micron$ diameter opaque metallic spot deposited at the center to mask the central defect \citep{vvc}. We used this coronagraph for the results presented in this paper. 

After the dichroic, the PIAA optics mounted in a wheel can be moved in or out to apodize the IR beam. At the focal plane, all the PMCs mentioned above sit in a wheel that can be adjusted in the x, y and z directions via motorized actuators. The on-axis starlight diffracted by the FPMs in a downstream re-imaged pupil plane encounters a pupil wheel, which sits at an angle of 6$\degree$ as shown in fig.~\ref{f:scexao}. This pupil wheel consists of the RLSs corresponding to each FPMs at the focal plane. These pupil masks are made by lithographing a layer of chrome on a fused silica disk of 1.5-mm thickness. Figure~\ref{f:scexao} shows an example of a RLS for the VVC coronagraph. The chrome, corresponding to the reflective surface in this image has a reflectivity of only 60\% in near infrared while the rest is being absorbed. 

The RLS at the pupil plane blocks the diffracted starlight rejected outside of the geometrical pupil. This unused masked starlight is reflected towards a Near Infrared (NIR) detector in a re-imaged focal plane for low-order wavefront sensing. This detector will be referred to as the LLOWFS camera throughout the paper. The nulled coronagraphic PSF is directed towards two different NIR imaging optics via a selection of beamsplitters that can select the spectral content and the amount of flux between the two optical paths. One relayed optical path is towards the high frame rate internal NIR imaging camera and another one is towards HiCIAO.
 
The LLOWFS camera and the internal NIR imaging camera are InGaAs CMOS detectors with a resolution of 320 $\times$ 256 pixels, a frame rate of up to 170 Hz and a read out noise of 140~e$^{-}$. They are used for the alignment of the coronagraphs as well as the testing and calibration of the low-order control loop either with the internal calibration source or directly on the sky. On the other hand, HiCIAO uses a HAWAII 2RG detector with a resolution of 2048 $\times$ 2048 pixels, a frame rate $\ll$~3 Hz and a read out noise of 15~-~30~e$^{-}$. HiCIAO is a facility science instrument we used to perform the differential imaging and to collect the post-coronagraphic data during the on-sky operations. The advantage of having both the internal NIR camera and the HiCIAO is that the former can be used to track the high temporal frequencies in the atmospheric turbulence while the latter is ideal for tracking the slow varying spatial frequency components with much better sensitivity.

The SCExAO instrument is developed with an ultimate goal of being rapidly adaptable to the future extremely large telescopes \citep{elt}. Further details of the SCExAO instrument and its future capabilities are beyond the scope of this paper and are described in detail in \cite{scexao}.
%
%
%          ** 2.2) DM on SCExAO**
%================================

\subsection{Deformable Mirror as a wavefront corrector and a turbulence generator} 
{
The DM of SCExAO cannot only be used to control the aberrations up to the highest spatial frequency of 22.5 $\lambda/D$ but also to inject phase errors to simulate a dynamical turbulence for laboratory tests. The phase maps injected on the DM are built using a simulated phase screen, which follows the Kolmogorov profile. This phase screen can also be filtered to mimic the effects of the low and high spatial frequencies under pre/post-AO corrections. The simulated turbulence can run in the background independently of the corrections injected on the DM by the wavefront control loops. The final command sent to the DM is then the sum of the injected turbulence and the calculated corrections. For the turbulence injection, we control different parameters: strength (amplitude in nm~RMS), wind speed (m/s) and an optional coefficient reducing the low-spatial frequencies to mimic the effect of the AO188 wavefront residuals. However, this simulation is limited by the spatial frequency of the DM, which is 22.5 cycles/aperture.}
%
%
%
%          **2.3) LLOWFS Operation on SCExAO**
%============================================

\subsection{LLOWFS operation on SCExAO}
\label{s:llowfs}

SCExAO has a dedicated low-order wavefront correction loop, which uses the measurement of the LLOWFS to calculate the control commands. The measured aberrations are compensated by actuating the DM by the following two methods:

\begin{itemize} 
\item{Direct interaction with the DM: The low-order wavefront corrections are sent directly to the DM. In this case, 35 Zernike modes in the laboratory and 10 Zernike modes on sky can be controlled thus far. The method and the results obtained are described in detail in section \ref{s:llowfsDM}.}
\item{Indirect interaction with the DM: The second avenue of communication is when the LLOWFS controls the piezo-driven tip-tilt mount of the dichroic, which separates visible and IR channels, to offset the zero-point of the PyWFS. With this configuration, the axis of the PyWFS is changed by moving the dichroic in tip-tilt with the corresponding amount of measured pointing residuals. This pointing shift in the visible channel is then compensated by the DM in closed-loop,  hence indirectly controlling the differential pointing errors in the IR channel. We demonstrate the concept and the first on-sky results with this preliminary setup known as the differential pointing system, in section 4.}
\end{itemize}    

The second approach of low-order wavefront control is the one that will be used in the final configuration of SCExAO during the scientific observations. Indeed, the different wavefront sensors on SCExAO use the same DM for the wavefront correction, therefore cannot run simultaneously as  separate units. Nevertheless, the first approach is still valid for coronagraphic ExAO designs that have a dedicated DM for the low-order correction.
%
%             ** 3) LLOWFS with DM**
%================================

\section{Low-order correction using direct interaction with the DM}
\label{s:llowfsDM}
%
%                   **3.1)  Config**
%================================

\subsection{Configuration}
\label{s:config1}
\begin{figure}[h]
   \centerline{
        \resizebox{0.5\textwidth}{!}{\includegraphics{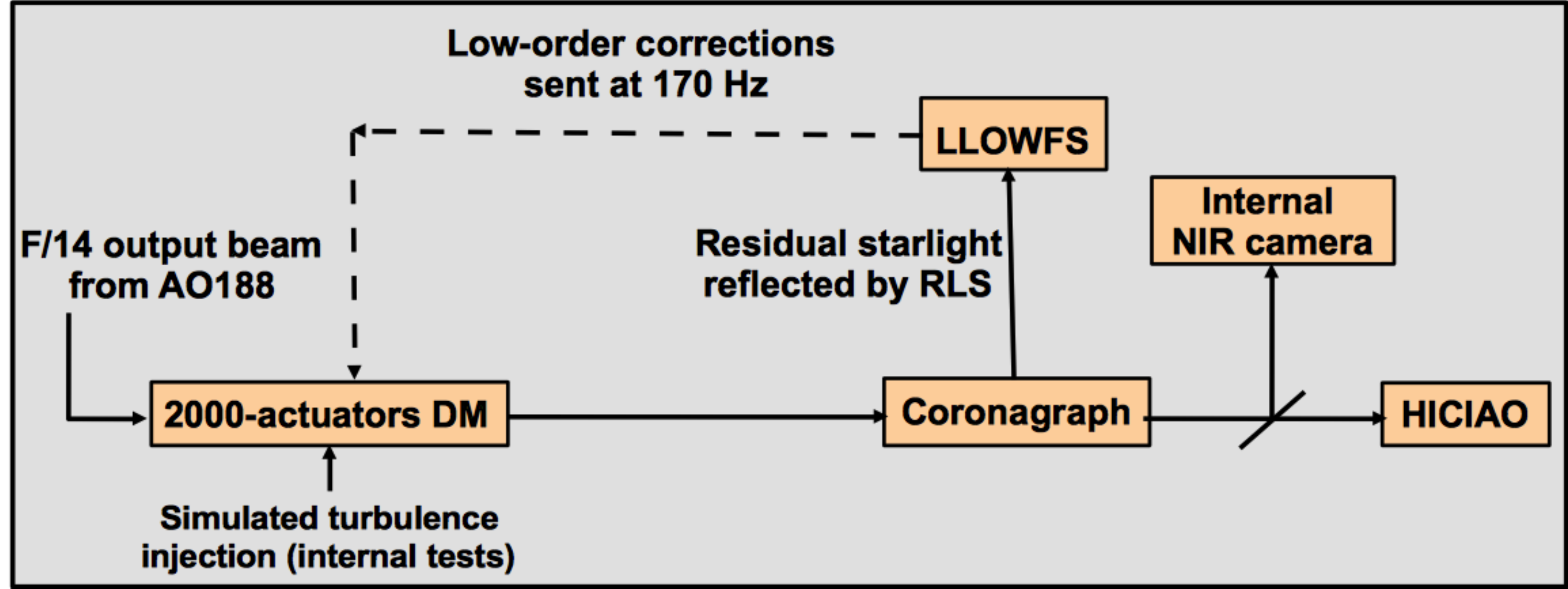}}}
   \caption{Flowchart of the configuration when the LLOWFS is directly coupled to the DM as the actuator on the IR channel of SCExAO. The LLOWFS camera senses the starlight reflected by the Lyot stop and measures the low-order aberrations. Calculated corrections are then sent to the DM. In this configuration, we use a simple integrator control law.}
  \label{f:lowfs1}
\end{figure}

Figure \ref{f:lowfs1} summarizes the configuration in a simplified flowchart. The starlight rejected by the coronagraph is reflected towards the LLOWFS camera. The reflected intensity at any instant is then sensed at the rate of 170~Hz. The low-order estimations are done by first obtaining the response matrix, also called calibration frames. These frames are acquired by applying a known amplitude of each Zernike mode independently to the DM. The reference subtracted response of the sensor is saved as a response matrix. The measurements are obtained using the Singular Value Decomposition (SVD) method, and used by an integrator controller to compute the corrections. These corrections are then sent to the DM, which compensates for the low-order aberrations. 

%
%
%    **3.2) Calibration**
%================================

\subsection{Calibration frames acquisition}
\begin{figure*}
   \centerline{
        \resizebox{0.9\textwidth}{!}{\includegraphics{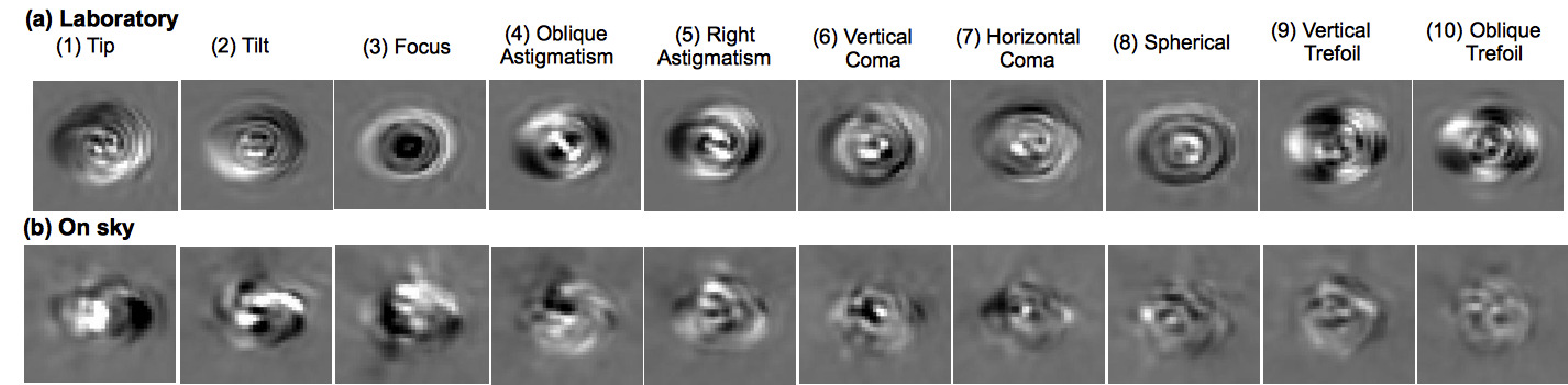}}}
   \caption{Response matrix for the VVC obtained (a) in the laboratory and, (b) on sky for 10 Zernike modes. Note: These frames have the same brightness scale.}
  \label{f:cal}
\end{figure*}

Figure \ref{f:cal} presents the response of the LLOWFS to probe the low-order Zernike modes. These frames are acquired prior to closing the control loop. In the laboratory, without any simulated turbulence, we apply a phasemap with an amplitude of 60~nm~RMS for the 35 Zernike modes separately to the DM. The effect of these modes on the low-order images is subtracted from the reference frame to calibrate the LLOWFS response to the low-order modes. Figure~\ref{f:cal}~(a)(1)-(10) shows the response matrix obtained in the laboratory for 10 Zernike modes only. This figure shows a clear distinction between the calibration frames, indicating no confusion in the response of the LLOWFS to different low-order modes. 

In a similar manner, Fig.~\ref{f:cal}~(b)(1)-(10) shows the on-sky calibration frames obtained by applying phasemaps with an amplitude of 60~nm~RMS for the 10 Zernike mode on the DM while observing the science target Epsilon Leonis (1.5~mas~RMS of tip-tilt angle on sky). These calibration frames were obtained with the AO188 loop closed. 

The on-sky response matrix looks noisier than the one obtained in the laboratory. It is actually dominated by uncorrected phase errors, since the AO188 is the only loop providing wavefront correction. Even if the on-sky signal is not as strong as in the laboratory, the modes are quasi-orthogonal and still can be used to close the loop.
%
%
%  **3.3) Measurements **
%================================

\subsection{Measurements}
In order to characterize the performance of a low-order wavefront sensor for coronagraphic purpose, it is important to understand how efficiently the pointing errors are measured and mitigated. We analyzed the properties like the linear response of the sensor, the cross coupling between the low-order modes and the requirement of how often the calibration frames should be reacquired. 
%
% **3.3.1) Linearity **
%================================

\subsubsection{Linearity}

\begin{figure}[h]
   \centerline{
        \resizebox{0.5\textwidth}{!}{\includegraphics{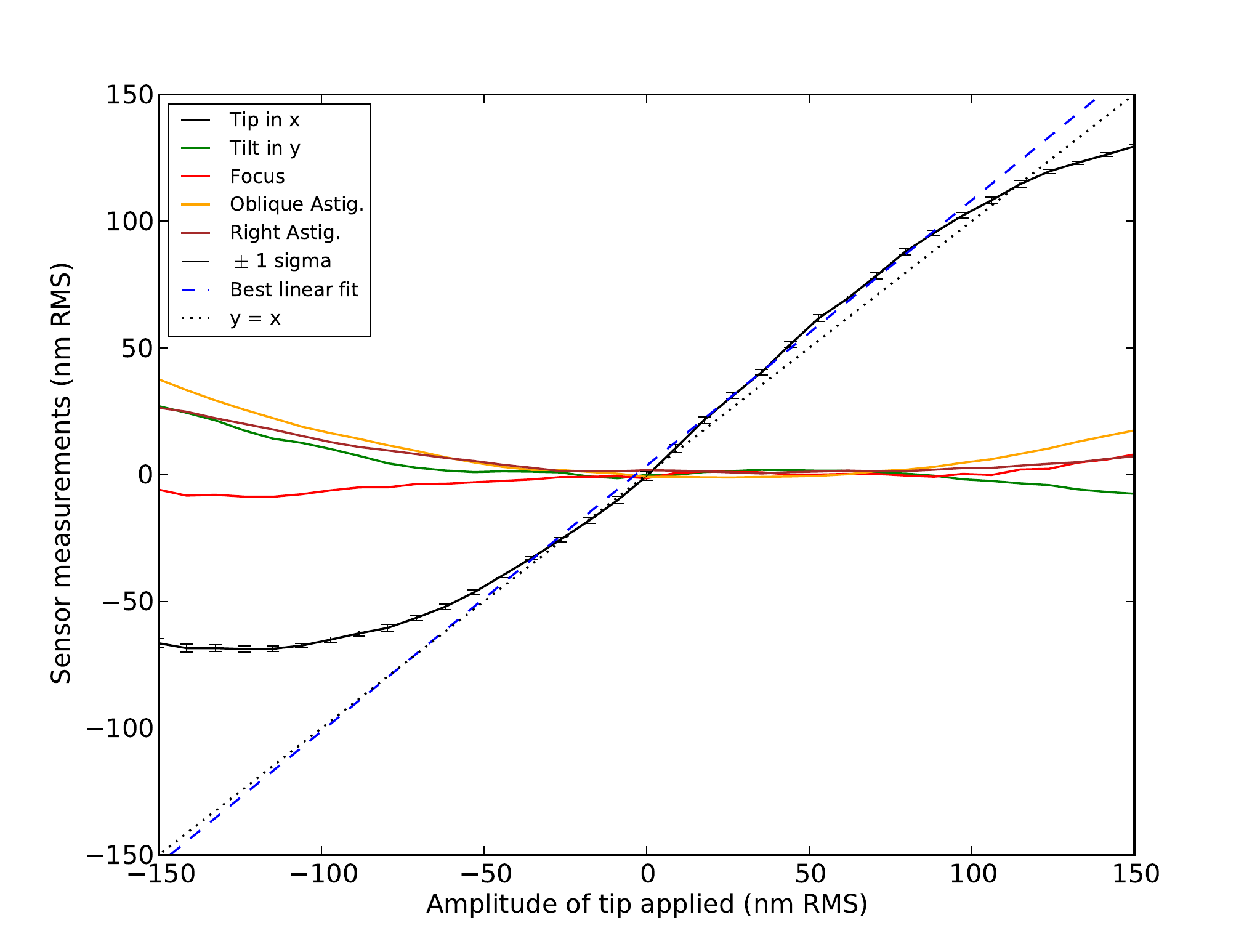}}}
   \caption{Linear response of the sensor to the tip aberrations in the case of the VVC. The Y-axis shows the measurements estimated for 5 modes. The residuals of tilt, focus, oblique and right astigmatisms are $\sim$~1~nm~RMS within the linear range. The blue dash line shows the best linear fit within the linear range (from -~50~nm to 100~nm~RMS) of the sensor. Note: The plotted data is the average of the aberrations estimated in a set of 20 measurements.}
  \label{f:flin}
\end{figure}

Figure \ref{f:flin} presents the linearity of the sensor to the tip aberration studied in case of the VVC. We applied phasemaps of tip aberrations with amplitudes between $\pm$~150~nm~RMS to the DM. The impact of each phasemap on the low-order images was recorded. Using the response matrix acquired in Fig.~\ref{f:cal} (a), the amount of the tip error as well as the residual in the other modes was estimated through SVD. The experiment was repeated 20 times and the plotted data is the average of the 20 measurements acquired. 

The linearity range of the sensor is around 150~nm~RMS (from -~50~nm to 100~nm~RMS) for the tip mode in x. The residuals of the modes tilt in y, focus, oblique and right astigmatism extracted through SVD are $\sim$~1~nm~RMS within the linearity range which is a tolerable amount of cross-coupling between the modes. The shift in the center of the linear range towards one direction could be caused by misalignments of the beam with respect to the FPM, or by the 25-$\micron$ metallic dot not being perfectly centered with the vortex half-waveplate. We repeated the linearity test with the rest of the modes and observed a similar behavior in the range of linearity and the shift of the zero point. 

Therefore, the stability of the reference image on the low-order camera dictates how often the LLOWFS should reacquire calibration frames. During the acquisition of the calibration, if the environmental factors, such as temperature variation and the flexure of the instruments, introduce tip-tilt errors in the reference PSF, then the system needs to be re-calibrated. If these drifts happen prior to closed-loop operation and are out of the linearity range, then only the PSF need to be realigned behind the FPM and previously acquired calibration frames can be reused to close the loop. However such drifts will not affect the closed-loop operation as the low-order correction will compensate for them.
%
%
% ** 3.3.2) Turbulence Injection **
%================================

\subsubsection{Turbulence injection in the Laboratory}

\begin{figure}[h]
   \centerline{
        \resizebox{0.4\textwidth}{!}{\includegraphics{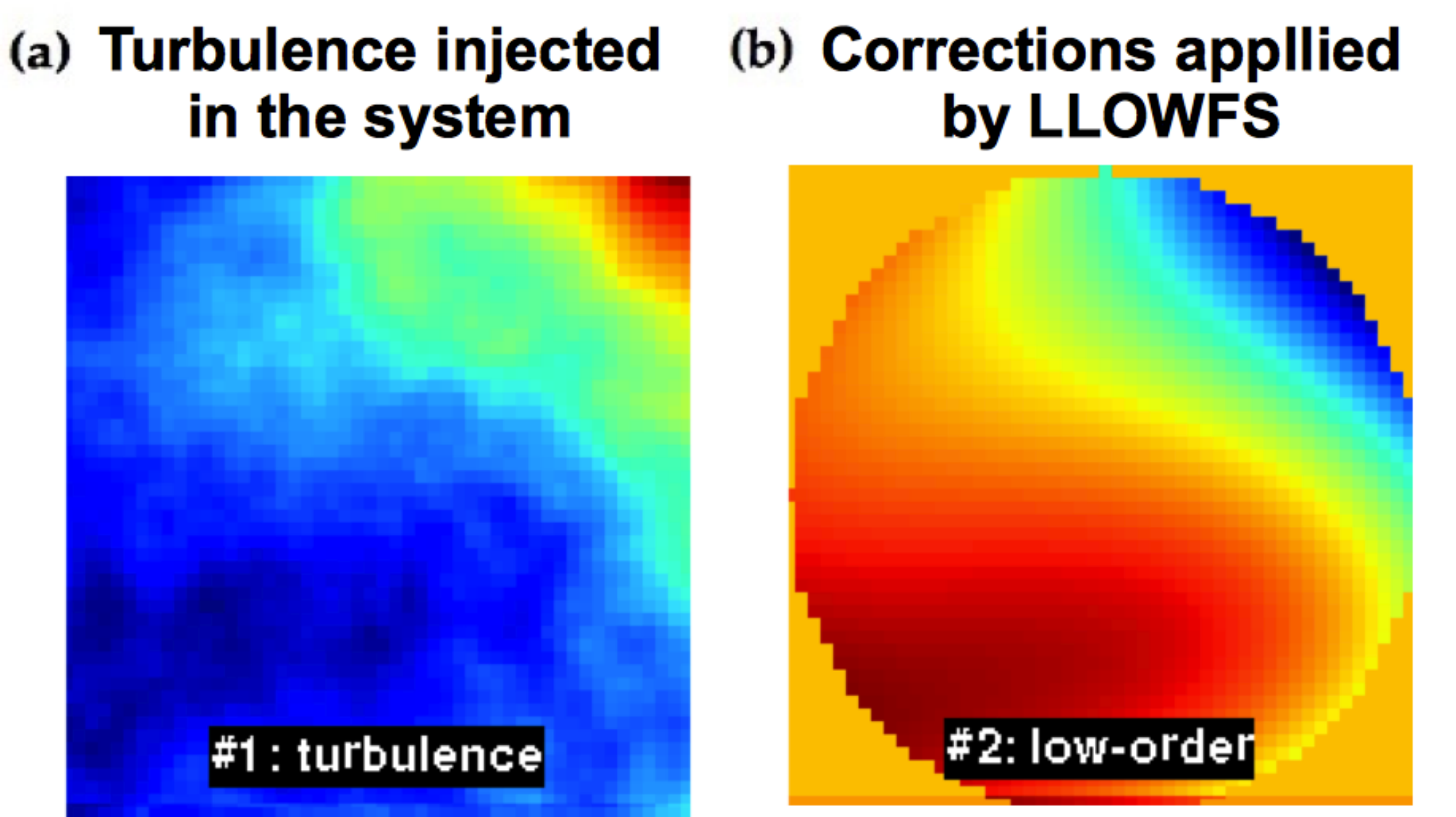}}}
   \caption{(a) The figure shows one phase map of the dynamic turbulence injected into the system (on the DM) and, (b) the corrections computed using the LLOWFS in the laboratory. During closed-loop operation of the LLOWFS, the final command that is being sent to the DM is the sum of these phasemaps.}
  \label{f:tur}
\end{figure}

All of our experiments in the laboratory are conducted with simulated dynamic phase errors that were applied on the DM. For the turbulence simulation, we chose 150~nm~RMS as the amplitude, 10~m/s as the wind speed and we allowed all the low-spatial frequency components of the turbulence to be left uncorrected mimicking the case with no AO correction upstream. Figure~\ref{f:tur} (a) is the visualization of a phasemap of the simulated turbulence applied on the DM. 

%
%
% **3.3.3) Spectral analysis in Lab
%================================

\subsubsection{Spectral analysis in the laboratory}
\label{s:lab}

For the laboratory test presented here, the low-order control loop is correcting 35 Zernike modes at 170~Hz,  the frequency of the camera. The gain of the integrator controller is set to 0.7. We can push the gain to high values because the latency of the control loop is very low, $\sim$~1.1~frames.   

Figure \ref{f:tur} (b) shows the correction phasemap computed by the LLOWFS control loop corresponding to the turbulence applied in Fig.~\ref{f:tur} (a). As expected, the color map in both images are opposite to each other, i.e.\@ the control command cancels the injected turbulence. In closed-loop operation, the final command applied to the DM is the sum of these two phasemaps.  

The frequency of the LLOWFS (170~Hz) is much higher than the maximum frequency resolved by the minimal exposure time of the science detector HiCIAO ($<$~0.5~Hz for an exposure time of 2~seconds). So to have a meaningful evaluation of the residuals in open and closed loop, we will analyze them in two temporal bands: 
\begin{itemize}
\item{0~-~0.5~Hz : corresponds to slow varying frequency components temporally resolved by the science camera, i.e.\@ the dynamical contribution of the turbulence in the science images of HiCIAO.}
\item{0.5~-~85~Hz : corresponds to the faster motions resolved by the LLOWFS but averaged by the exposure time of the science camera, i.e.\@ the static contribution of the turbulence and the vibrations in the science images. }
\end{itemize} 

\begin{figure*}
   \centerline{
        \resizebox{1.0\textwidth}{!}{\includegraphics{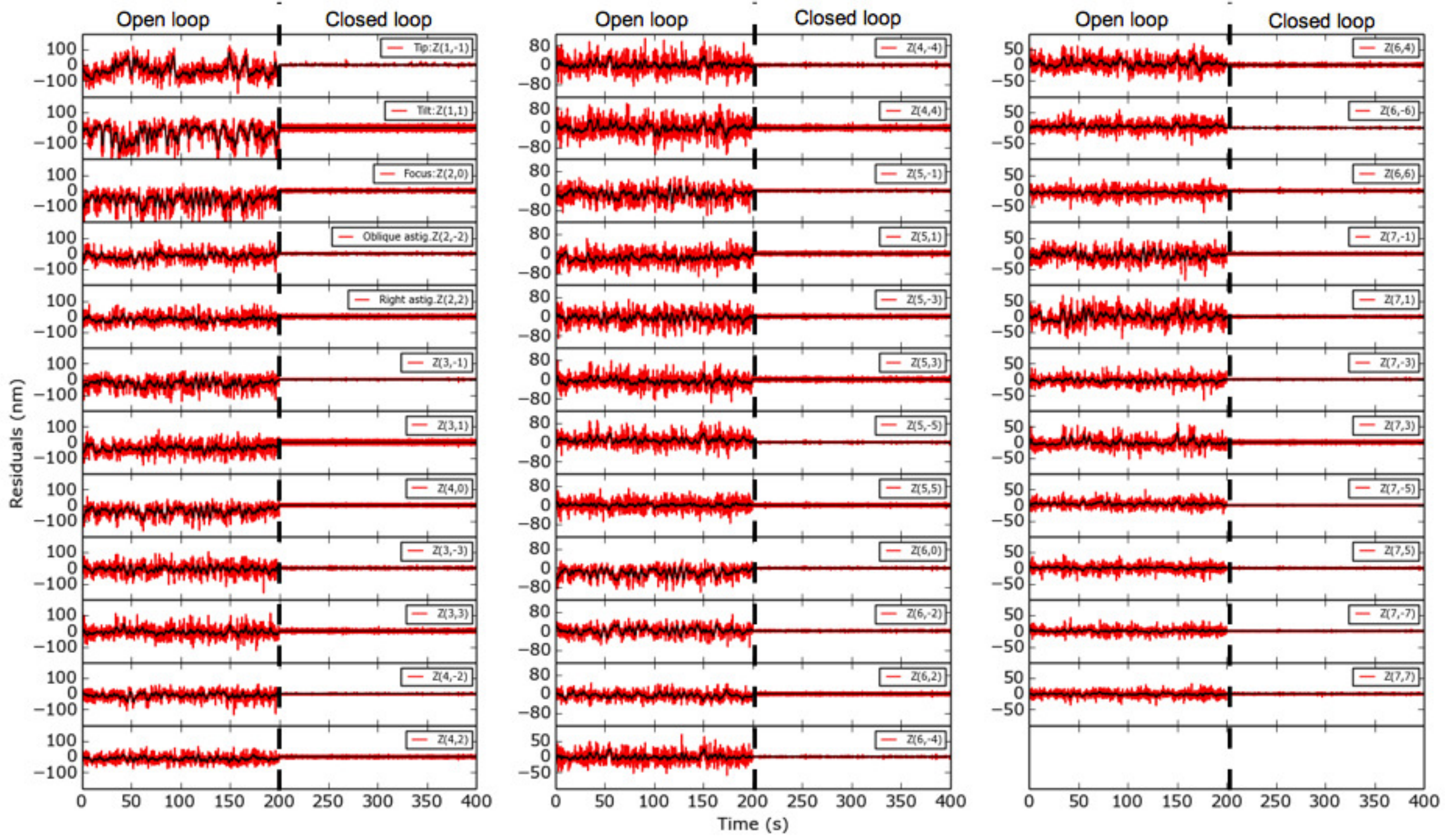}}}
   \caption{Residuals in open and closed loop for 35 Zernike modes obtained in the laboratory with dynamic turbulence. The red lines are the raw residuals while the black lines are the moving average of the residuals using a 2-second window. Figure~\ref{f:stdLab} quantifies the open- and closed-loop residuals for the measurements presented here.}
  \label{f:resLab}
\end{figure*}

Figure~\ref{f:resLab} presents a temporal measurement of the open- and closed-loop residuals for 35 Zernike modes. These measurements (red lines) are filtered by a moving average of 2~seconds to match the minimal exposure time of HiCIAO (black lines). In closed-loop operations, the stability of the residuals improved noticeably for all the modes. 

 \begin{figure}[h]
   \centerline{
        \resizebox{0.5\textwidth}{!}{\includegraphics{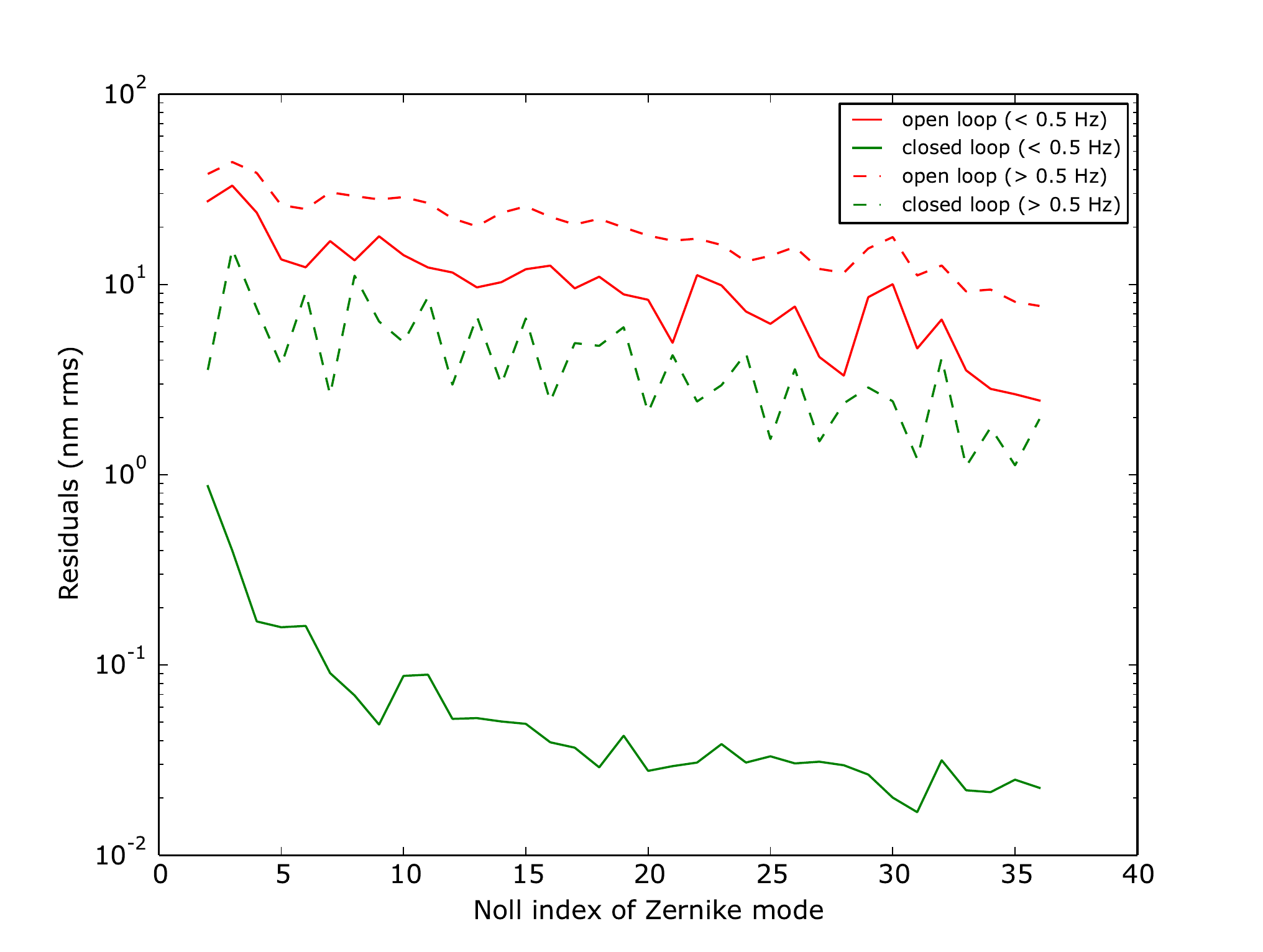}}}
   \caption{Open- and closed-loop residuals for 35 Zernike modes corrected under the laboratory turbulence. The correction at low frequencies is about two orders of magnitude, leaving sub-nanometer residuals for all the modes. }
  \label{f:stdLab}
\end{figure}

Figure~\ref{f:stdLab} summarizes the open- and closed-loop residuals for all 35 Zernike modes. We obtained a reduction by a factor of 30 to 500 (median of 200) on all the modes for the low frequencies ($<$~0.5~Hz), leaving only sub-nanometer residuals. For the higher frequencies ($>$~0.5~Hz), the factor of improvement is only between 3 and 12 (median of 5), because it is dominated by the vibrations that are not corrected by the controller. These vibrations, mostly coming from the resonance at 60~Hz of a Stirling cooler, are introduced by mechanical motions of the optical elements on the bench. In fact the vibrations above 10~Hz are actually amplified by the overshoot of the controller. The pointing residuals for open- and closed-loop sampled at 0.5~Hz are about $10^{-2}$~$\lambda/D$~RMS (0.8~mas) and a few $10^{-4}$~$\lambda/D$~RMS (0.02~mas) respectively. 

 \begin{figure}[ht]
   \centerline{
        \resizebox{0.5\textwidth}{!}{\includegraphics{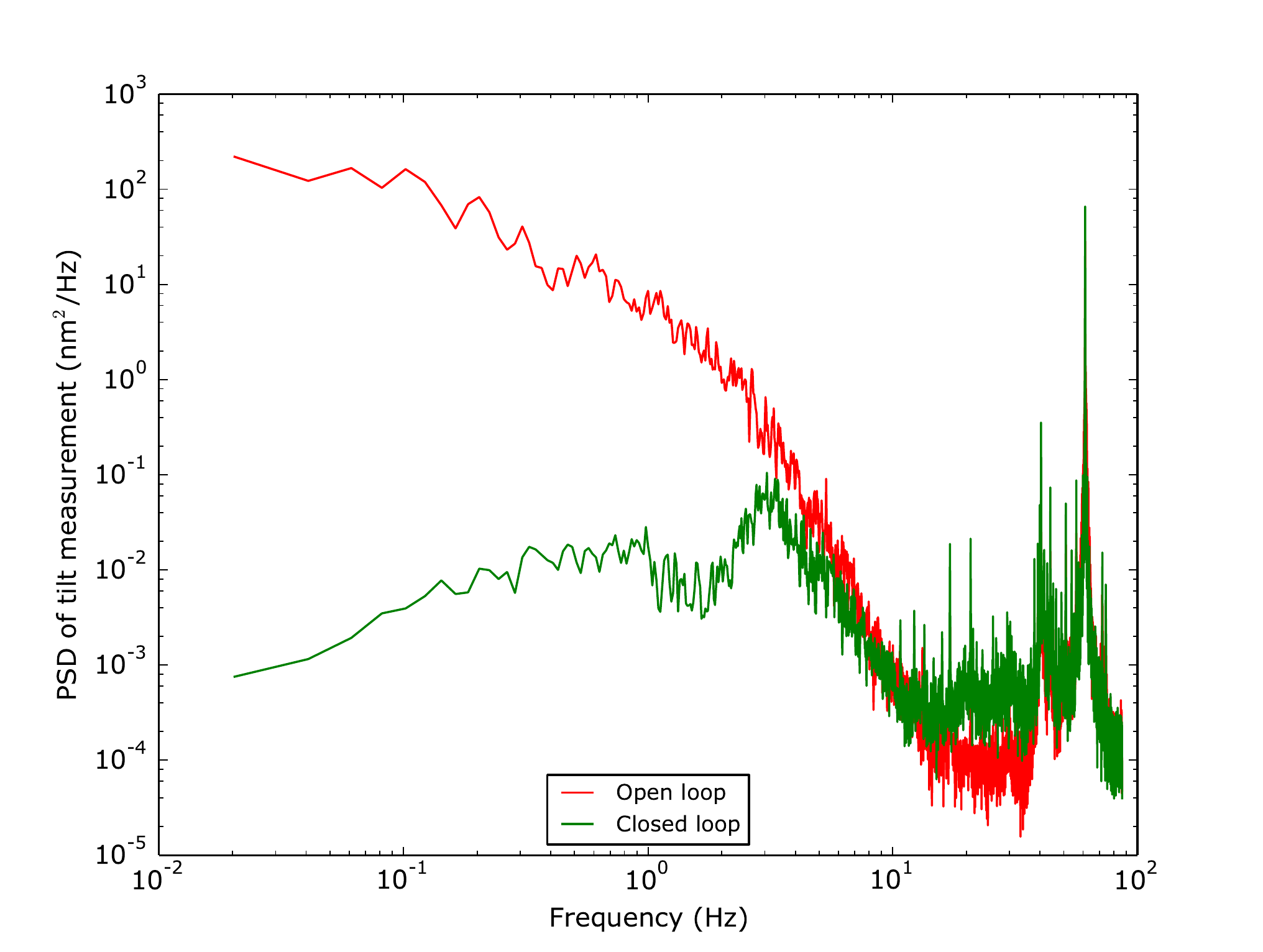}}}
   \caption{PSD of the open- and closed-loop for the tilt aberration under the laboratory turbulence. Significant improvement is visible in closed loop operation at frequencies $<$~3~Hz. The vibrations beyond 10~Hz are amplified due to the overshoot of the controller.}
  \label{f:psdLab}
\end{figure}

The high speed of the LLOWFS helps us to analyze the vibrations induced either by mechanical (cryo-coolers, motors etc.\@) or environmental (telescope structure due to wind-shaking) factors. In order to analyze the spectral distribution, we study the Power Spectral Density (PSD) of the residuals. The PSD is calculated as the square modulus of the Fourier transform of the residuals. A Welch smoothing is performed on the PSD to reduce the noise. Figure \ref{f:psdLab} presents the PSDs of the open- and closed-loop data of only the tilt mode in the laboratory. The improvement is about two orders of magnitude at 0.5~Hz while high frequencies $>$~10~Hz are slightly amplified. 

We have yet to identify the source of vibrations occurring beyond 10~Hz, which are probably due to optical elements vibrating inside the instrument. These oscillations are for now beyond the bandwidth of the low-order wavefront controller and therefore amplified by its overshoot. We are currently optimizing this control loop with a LQG controller to correct for the vibrations of the telescope and the instrument. The LQG, based on a Kalman filter, uses the real-time low-order telemetry to calculate a model of the disturbance (pointing errors, turbulence and vibrations) and predicts the best correction to apply. Further discussions of LQG implementation on SCExAO will be the focus of a future publication.
%
%
% ** 3.3.4) Spectral analysis on-sky
%================================

\subsubsection{Spectral analysis on-sky}

After having tested the LLOWFS in the laboratory conditions, we analyzed its performance during an on-sky engineering run in April 2015. The results presented here were taken on the science target Epsilon Leonis ($m_{H} = 1.23$). In this case, AO188 closed the loop on 187 modes providing a Strehl ratio of $\sim$~40\% (500~nm~RMS wavefront error) in H band. The LLOWFS then closed the loop on this wavefront residuals at 170~Hz with 10 Zernike modes. Since the gain of the loop is tuned manually at present, a conservative gain of 0.05 is used for this demonstration to ensure the stability of the closed-loop operation.   

\begin{figure*}
   \centerline{
        \resizebox{0.8\textwidth}{!}{\includegraphics{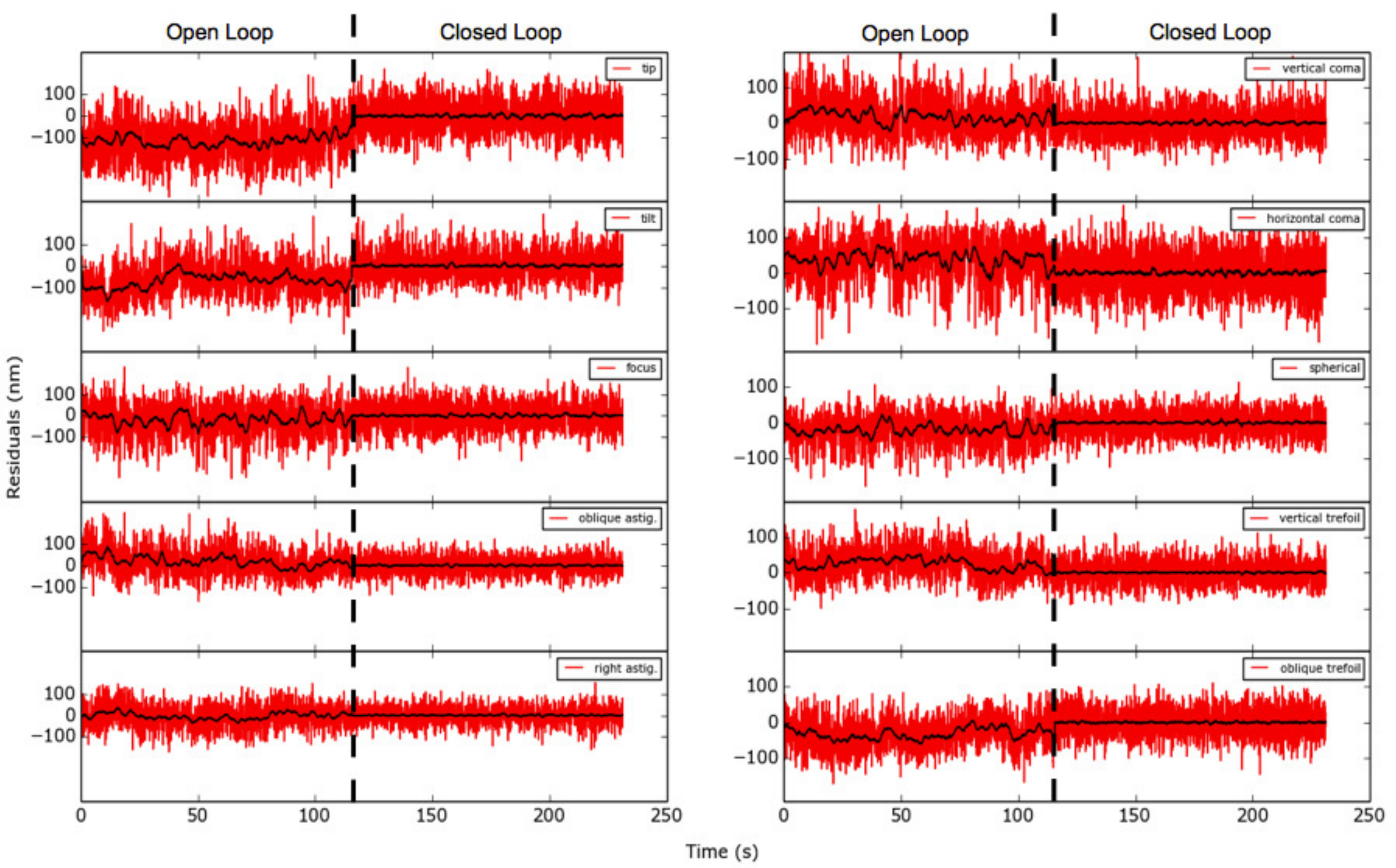}}}
   \caption{On-sky open and closed loop residuals for 10 Zernike modes for the science target Epsilon Leonis. The red lines are the raw residuals whereas the black lines are the moving average with a  2-second window. Figure~\ref{f:stdOnsky} quantifies the residuals presented here. Note: The open loop is the post-AO188 raw residuals and the amplitude variations of the residuals are sometimes outside of the linear range of LLOWFS, which cause the underestimation of their measurement.}
  \label{f:onskyRes}
\end{figure*}

Figure~\ref{f:onskyRes} presents the on-sky open- and closed-loop residuals. Similarly to Fig.~\ref{f:resLab}, the results are smoothed by a moving average using a window of 2~seconds to match the minimal exposure time of HiCIAO. The improvement in the closed-loop residuals is visible in the on-sky data. However, the residuals are more disturbed by vibrations, hence are noisier than those collected in the laboratory. 

\begin{figure}[h]
   \centerline{
        \resizebox{0.5\textwidth}{!}{\includegraphics{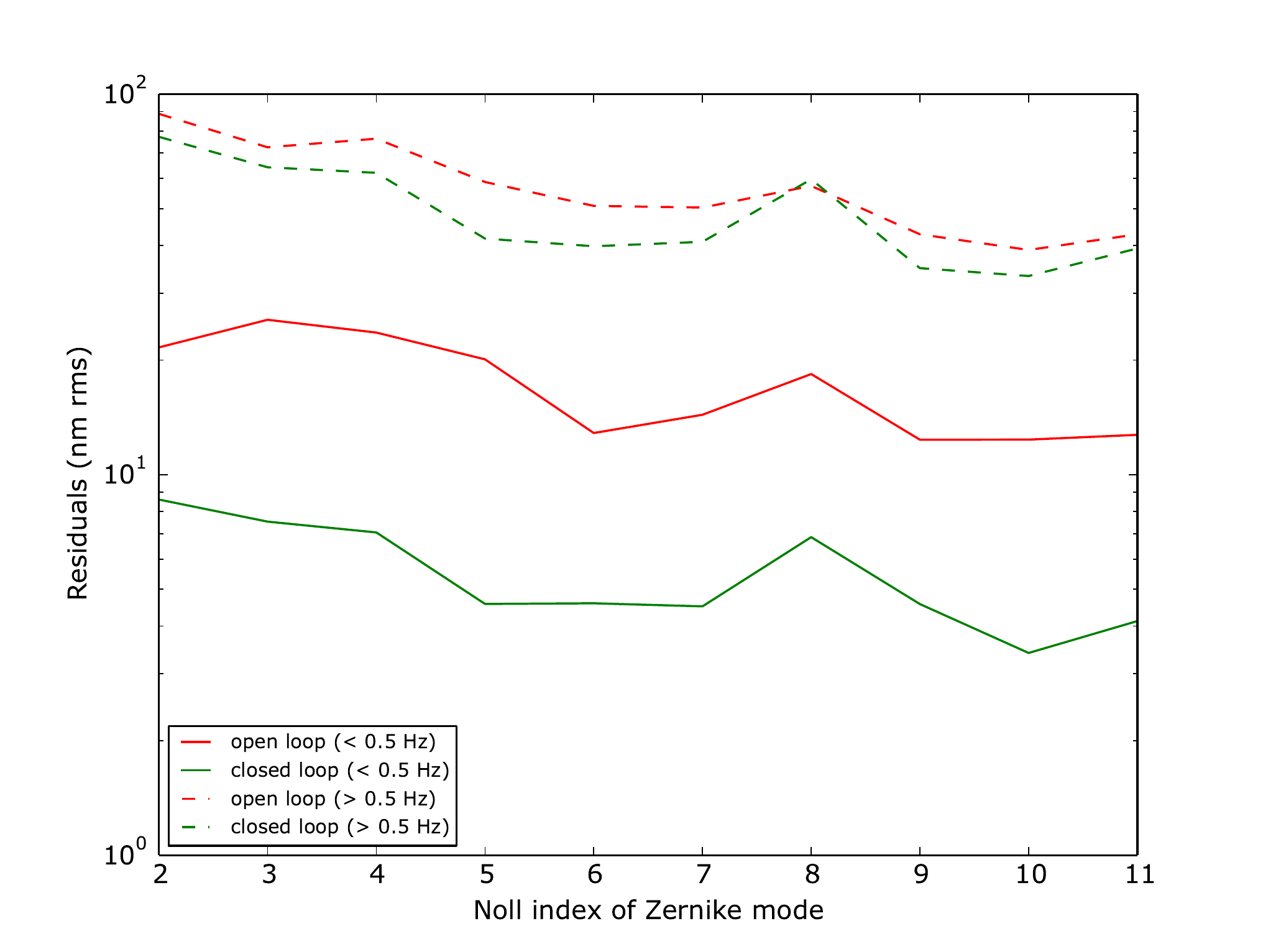}}}
   \caption{Open- and closed-loop residuals obtained on-sky for 10 Zernike modes. The correction provides a significant improvement at low frequencies but slightly amplifies the higher frequencies.}
  \label{f:stdOnsky}
\end{figure}

The same analysis as the one explained in Sec.~\ref{s:lab}, i.e. separating low frequencies below 0.5~Hz resolved by HiCIAO and the high frequencies above 0.5~Hz averaged by HiCIAO, was performed on the on-sky data and is presented in Fig.~\ref{f:stdOnsky}. For low frequencies, we obtained a reduction by a factor of 2.5 to 4.4 (median of 3.1) for all the modes while for the higher frequencies, closing the loop corrected the residuals by a factor of 1.2 only. This is expected due to the small gain value of the integrator controller. However, we demonstrate that the slow varying pointing errors are reduced down to a few $10^{-3}$~$\lambda/D$~RMS (0.15~mas). 

 \begin{figure}[h]
   \centerline{
        \resizebox{0.5\textwidth}{!}{\includegraphics{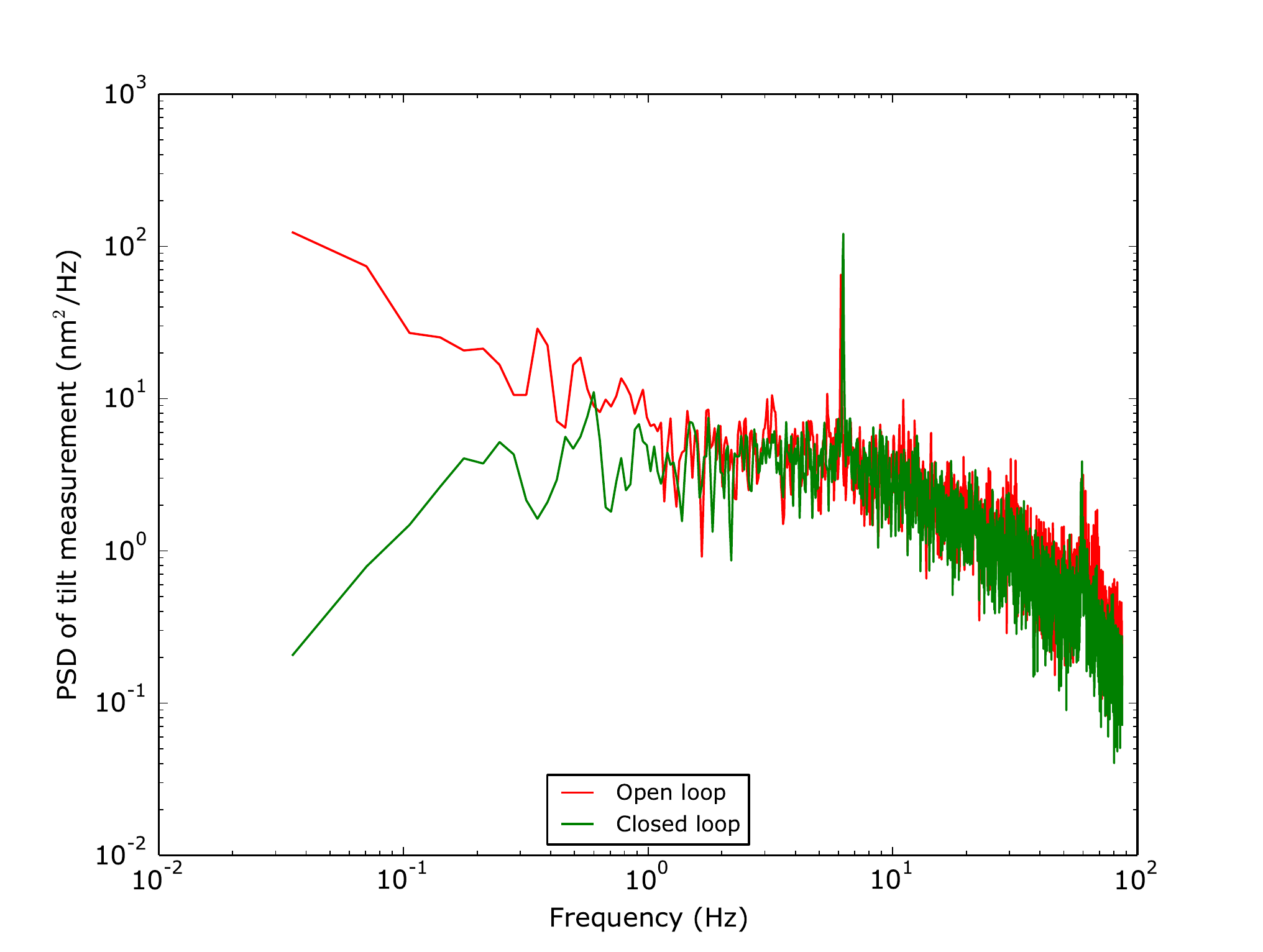}}}
   \caption{On-sky PSD of the open- and closed-loop presented for the tilt aberration only. A telescope vibration around 6~Hz appeared during the on-sky operation. In closed-loop, an improvement can be noticed at frequencies $<$~0.5~Hz. Due to the effects of the non-linearities in LLOWFS response, the real amplitude of the residuals are underestimated, causing the slope of the PSD to appear smaller than the one obtained in the laboratory.}
  \label{f:onskyPSD}
\end{figure}

In Fig.~\ref{f:onskyPSD}, we present the on-sky PSD for the open and closed loop for the tilt aberration only. The profile of the disturbance is different from the laboratory experiment presented in Fig.~\ref{f:psdLab}. A new vibration around 6~Hz due to the telescope structure appeared in the on-sky PSD. The vibration at 60~Hz was reduced because the Stirling cooler causing it was removed from the instrument. Moreover, the shape of the pointing errors is different from the turbulence generated in the laboratory. Indeed, the general slope of the PSD is smaller than a typical Kolmogorov distribution. The amplitude of the variations are sometimes larger than the linear range of the LLOWFS ($\pm$~170~nm~RMS on the wavefront), which causes an underestimation of the real amplitude and a modification of the shape of the PSD. Due to a smaller gain value, the LLOWFS could not correct for the vibrations occurring beyond 0.5~Hz but the PSD shows a significant improvement below that frequency. Fig.~\ref{f:onskyPSD} summarizes the residuals in open- and closed-loop. However, because of the amplitude of the variations outside of the linear range, the values in the figure are probably underestimated.
%
%
%     **3.4) Processed Frames **
%================================

\subsection{Processed science frames}
\begin{figure*}
   \centerline{
        \resizebox{1.0\textwidth}{!}{\includegraphics{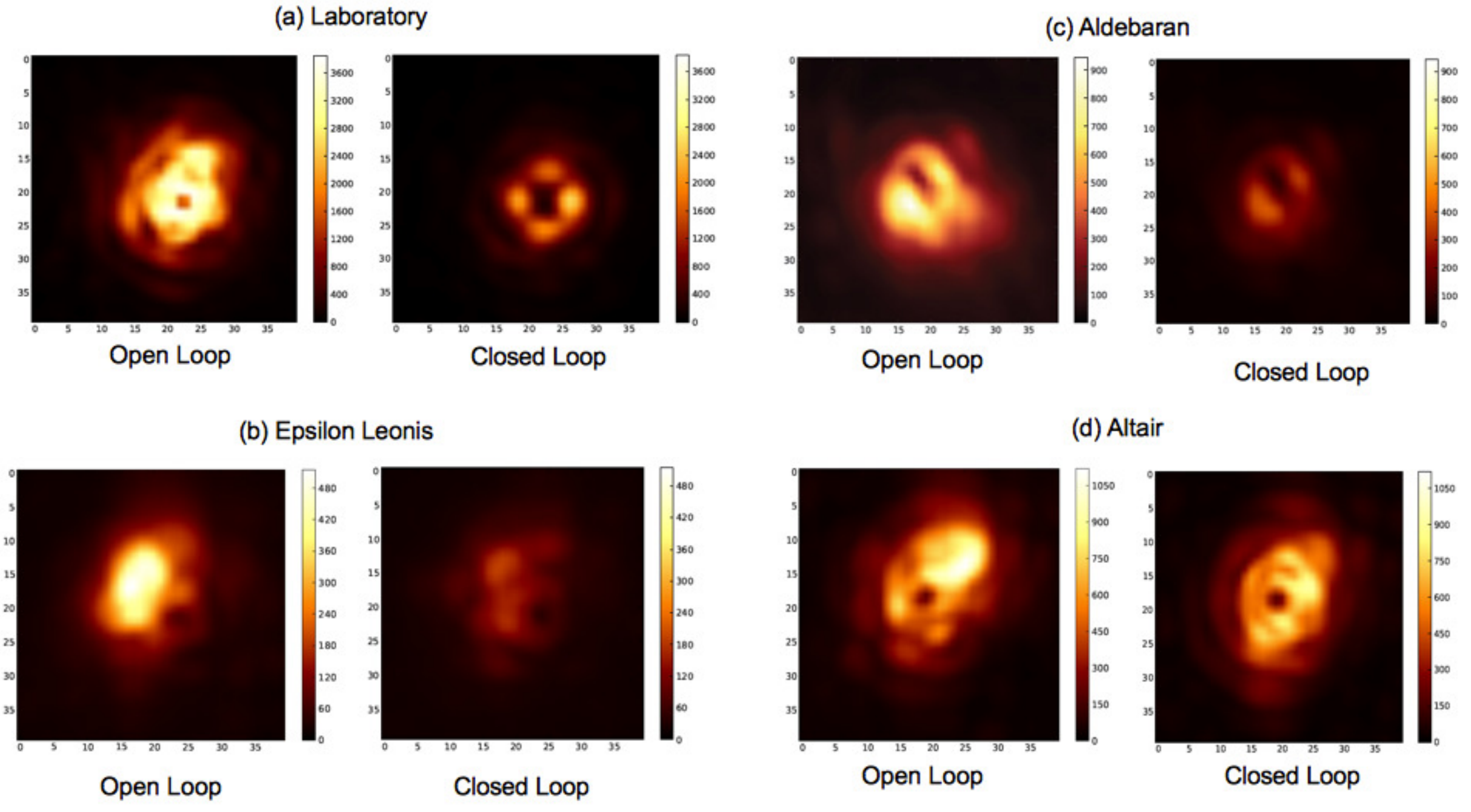}}}
   \caption{Comparison of the standard deviation of the intensity for 1000 frames of the NIR camera (a) laboratory, (b) Science target Epsilon Leonis ($m_H = 1.23$), c) Science target Aldebaran ($m_H = -2.78$) and the (d) science target Altair ($m_H = 0.10)$. Note: Each set of open- and closed-loop images are of same brightness scale. Closed-loop images are expected to be darker than the open loop images. Black spot at the middle of all the frames is the metallic dot at the center of the VVC to mask its central defects. }
  \label{f:frames}
\end{figure*}

In this section, we present the impact of the tip-tilt and other low-order residuals on the frames acquired by the SCExAO's internal NIR camera. Figure \ref{f:frames} presents the standard deviation per pixel in a cube of 1000 science frames (2~ms of integration time) for open and closed-loop, in the laboratory (Fig.~\ref{f:frames}~(a)) and on-sky for Epsilon Leonis (Fig.~\ref{f:frames}~(b)), Aldebaran (Fig.~\ref{f:frames}~(c)) and Altair (Fig.~\ref{f:frames}~(d)). These images show lower standard deviation for close loop images (hence darker than the open loop images) and a better centered beam behind the VVC in closed loop. However for the target Epsilon Leonis, the coronagraph was not centered perfectly when the reference frame was acquired. 

These images were obtained without the correction of high-order modes by the PyWFS. The LLOWFS in closed-loop only stabilizes the beam upstream of the VVC, without showing any significant contrast improvement in the absence of an ExAO loop. Therefore, the on-sky contrast enhancement of the VVC cannot be evaluated with these results. 

 \begin{figure*}
   \centerline{
        \resizebox{0.8\textwidth}{!}{\includegraphics{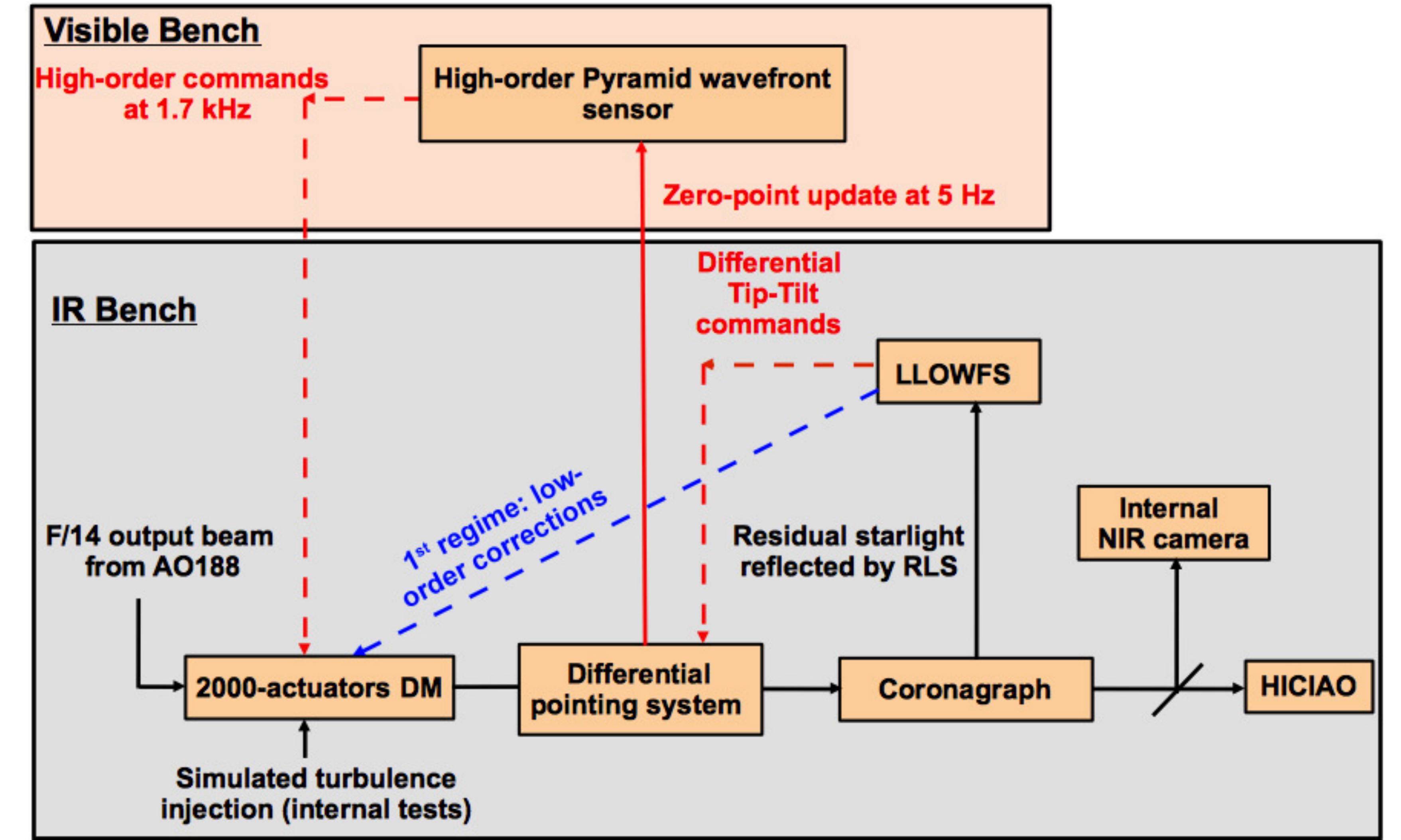}}}
   \caption{Flowchart of the LLOWFS functioning in two configurations on SCExAO. The black arrows depicts the common flow of the low-order control loop in both regimes. Configuration 1 (blue arrow) is when the LLOWFS is used directly with the DM to correct for the low-order aberrations as presented in section~\ref{s:config1}. Configuration 2 (red arrows) is when LLOWFS, after sensing differential tip-tilt errors in IR channel, updates the zero-point of the PyWFS using a differential pointing system. To compensate for the beam shift in the visible channel, the high-order loop commands the DM to correct for the chromatic errors.}
  \label{f:lowfs2}
\end{figure*}
%
%
%
%   **4) LLOWFS with Pyramid
%================================

\section{LLOWFS integration with the High-order Pyramid wavefront sensor}
\label{s:pyr}

The final goal of the LLOWFS is to work in close interaction with a high-order wavefront sensor like PyWFS to correct for the non-common path and chromatic errors occurring between the imaging and wavefront sensing channels. The control of non-common path aberrations is essential because the PyWFS is using the visible light while the coronagraph uses the NIR light. Also the PyWFS is not sensitive enough to low-order modes, and leaves a part of them uncorrected. So these uncorrected aberrations (static and dynamic) create unwanted stellar leakage around the coronagraphic mask in NIR. We integrated the LLOWFS with PyWFS to address these non-common path and chromatic errors. 
%
%
% **4.1) Config **
%================================

\subsection{Configuration}

SCExAO's high-order PyWFS, currently under development, is capable of controlling $\sim$~1600~modes at 3.6~kHz. For the results presented in this paper, we used an earlier version of the PyWFS running at 1.7~kHz and correcting only tip-tilt.

In this preliminary setup, PyWFS is the only system communicating with the DM. So instead of sending commands to the DM, the LLOWFS uses the differential pointing system to offset the zero-point of the PyWFS. Figure~\ref{f:lowfs2} presents the flowchart of the LLOWFS integration inside the high-order control loop. The blue arrow shows the first regime described earlier in Sec.~\ref{s:config1} (cf. Fig.~\ref{f:lowfs1}), where LLOWFS sends commands directly to the DM. The red arrows describes the configuration of the second regime where the high-order loop corrects for tip-tilt aberrations in visible and the low-order loop send commands to the differential pointing system to compensate for chromatic errors in IR. 

%
%
%   *4.2) On-sky demonstration
%================================

\subsection{On-sky demonstration}
We pointed the telescope to the variable star $\chi$ Cyg ($m_H =  -1.1$ during this observation). AO188 closed the loop on 187 modes with a seeing of 0.8" at 1.6~$\micron$. The PyWFS closed its loop only on tip-tilt in the visible with a 1.7~kHz loop speed. The PyWFS was not optimized at this point and hence provided only a partial correction of tip-tilt modes. 

\begin{figure}[h]
   \centerline{
        \resizebox{0.5\textwidth}{!}{\includegraphics{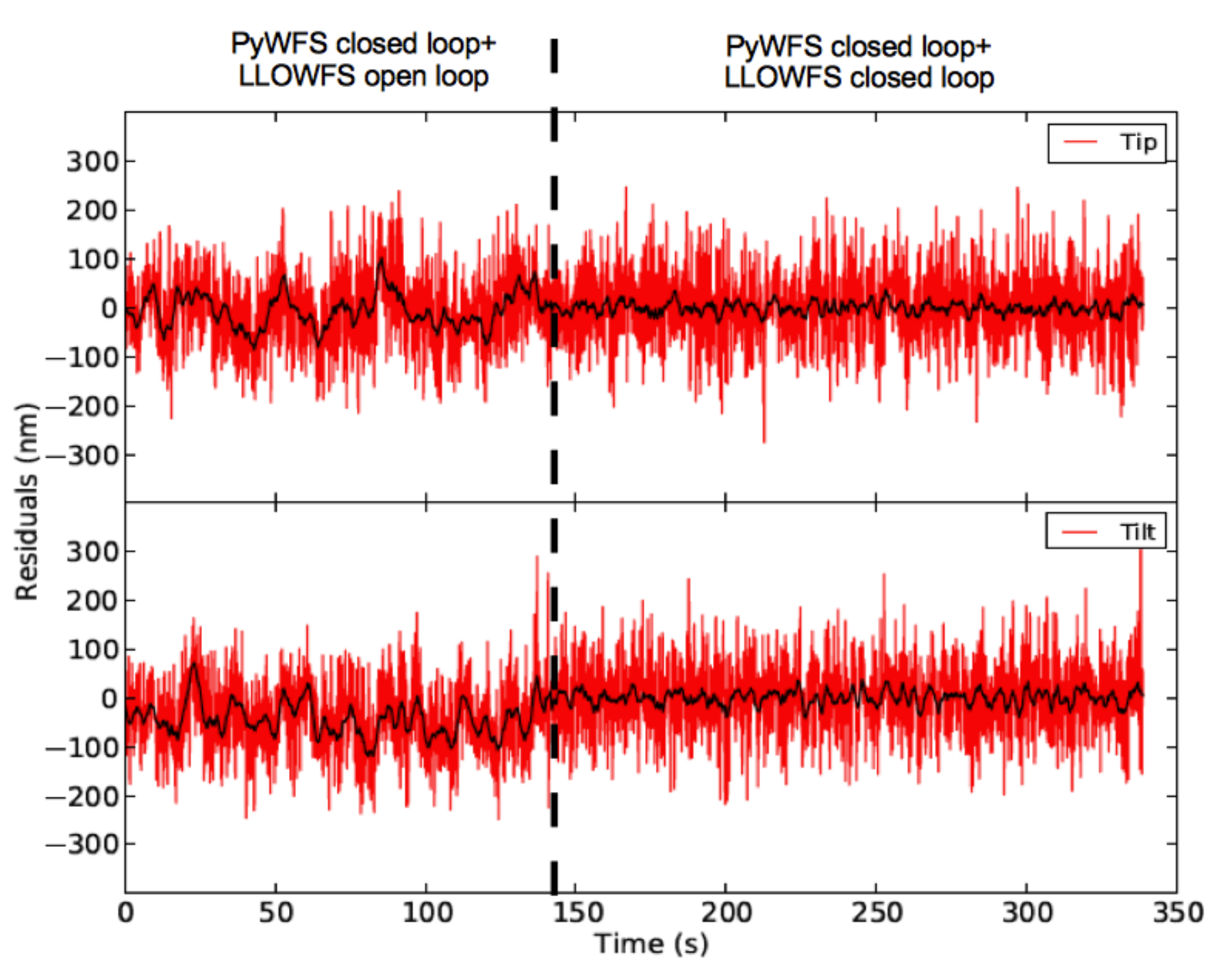}}}
   \caption{On-sky open- and closed-loop residuals of low-order control integrated in the high-order corrections of post-AO188 wavefront residuals. The black data is the moving average of residuals with 2~second window while the red data are the raw residuals. When the low-order loop is open then the high-order loop is correcting the pointing errors only in the visible leaving chromatic errors uncorrected. These chromatic errors are significantly reduced when the loop is also closed using the LLOWFS. Table~\ref{t:t3} summarizes low-order residuals for the differential tip-tilt modes. (Science target: $\chi$~Cyg.) }
  \label{f:resPyr}
\end{figure}

Similar to the first configuration explained in Sec.~\ref{s:config1}, the LLOWFS first acquired a response matrix in order to measure the non-common path errors. The on-sky reference is moved in x and y with an angle of 1.5~mas to obtain the calibration frames for differential tip and tilt. Using this response matrix, we closed the LLOWFS loop with a gain value of 0.03. A small gain was used because of the slow response of the piezos driver, controlled only up to 5~Hz. Figure~\ref{f:resPyr} shows the successful loop closure of the PyWFS and the LLOWFS. Once again, the data presented here are smoothed to simulate an exposure time of 2~seconds. When PyWFS loop is closed, we see a slight improvement in the stability, but a significant amount of non-common path residuals are still visible. These differential errors are improved when low-order loop is closed. 

Table~\ref{t:t3} summarizes the open and closed-loop residuals for high and low-order control loops. Once again, we analyzed the data at two different spectral bands. This table shows that we have achieved a factor 3 to 4 improvement in correcting differential tip-tilt residuals with the gain of 0.03 for the slow varying frequencies. As expected, the improvement for the higher frequencies is not significant due to the small gain used.

Due to the fact that the variations are larger than the linear range of the LLOWFS, the residuals in table~\ref{t:t3} are probably underestimated again. Even in such circumstances, closed-loop pointing residuals are only about $6\times10^{-3}~\lambda/D$ (0.23~mas) when the dataset is sampled at the frame rate of the science camera (0.5~Hz). 

\begin{figure}[h]
   \centerline{
        \resizebox{0.5\textwidth}{!}{\includegraphics{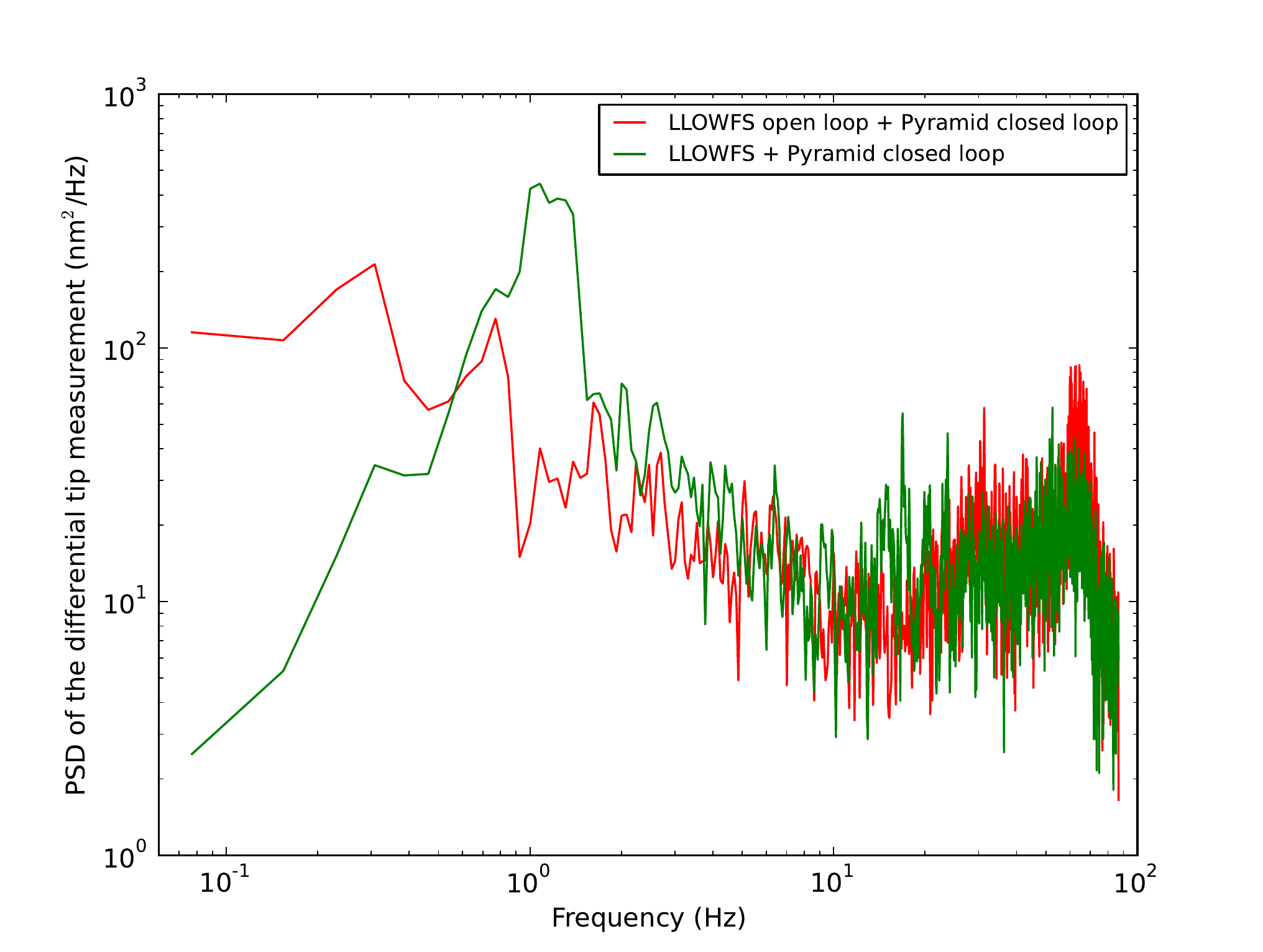}}}
   \caption{On-sky open- and closed-loop PSD of the differential tip aberration in case of the PyWFS integration with the LLOWFS. Closing the loop with the PyWFS reduces the telescope vibrations at  6~Hz shown in Fig.~\ref{f:onskyPSD}. The low-order correction provides significant improvement at frequencies $<$0.5~Hz and an overshoot around 1~Hz because of the difference in the sensing (170~Hz) and the correction (5~Hz) frequency.}
  \label{f:psdPyr}
\end{figure}

We present the on-sky PSD of the high and low-order integrated control loops for the differential tip aberration in Fig.~\ref{f:psdPyr}. Compared to the Fig.~\ref{f:onskyPSD}, the fast high-order control loop has diminished the telescope vibrations previously noticed at 6 Hz.  When we close the loop using the LLOWFS, we observe a significant reduction of the residual turbulence for low frequencies ($< 0.5~Hz$). In closed loop, an overshoot between 1 and 2~Hz is also visible. This is due to the mismatch between the frequency of the sensor (170~Hz) and the frequency of the actuator (5~Hz). A gain of 0.03 actually corresponds to a gain of 1 at the speed of the actuator, which explains the overshoot.

%
%          **  TABLE 1 **
%
\begin{table*}
\centering
\begin{tabular}{lccccc}
     &             & \multicolumn{2}{c}{\textbf{Low Freq. ($<$~0.5~Hz)}} & \multicolumn{2}{c}{\textbf{High Freq. ($>$~0.5~Hz)}} \\
     &             & \multicolumn{2}{c}{\textbf{(Resolved in HiCIAO)}}                & \multicolumn{2}{c}{\textbf{(Averaged in HiCIAO)}}    \\
     &             &                    &                      &                     &                       \\ \hline
Mode & Unit        & \textbf{Open loop} & \textbf{Closed loop} & \textbf{Open loop}  & \textbf{Closed loop}  \\ \hline
     &             &                    &                      &                     &                       \\
     & nm          & 26.1               & 9.4                  & 144                 & 142                   \\
Tip  & $\lambda/D$ & $1.6\times10^{-2}$ & $5.9\times10^{-3}$   & $9.0\times10^{-2}$  & $8.8\times10^{-2}$    \\
     & mas         & 0.66               & 0.24                 & 3.6                 & 3.6                   \\
     &             &                    &                      &                     &                       \\
     & nm          & 36.3               & 9.3                  & 170                 & 166                   \\
Tilt & $\lambda/D$ & $2.3\times10^{-2}$ & $5.8\times10^{-3}$   & $10.6\times10^{-2}$ & $10.4\times10^{-2}$   \\
     & mas         & 0.91               & 0.23                 & 4.3                 & 4.2                   \\                                                                                               
\end{tabular}
\caption\small{On-sky open- and closed-loop residuals of differential tip-tilt with the low-order loop integrated with the high-order loop. The correction is only significant for low frequencies.}
\label{t:t3}
\end{table*}

%
%    ** 4.3) Limitations
%================================

\subsection{Limitations with the initial setup}
The performance of the LLOWFS with its preliminary integration with the PyWFS was constrained due to several factors. 
\begin{itemize}
\item{Due to the slow response of the piezo driver (every 0.2 seconds), the LLOWFS could not control tip-tilt aberrations faster than 1~Hz. In the current configuration, we updated the differential pointing system by replacing the control of the tip-tilt from the piezo-driven dichroic to a tip-tilt mirror which is used for the modulation of the PyWFS. This will increase the loop rate up to 100~Hz.}
\item{Using either the dichroic or the tip-tilt mirror, the low-order control is limited to only tip and tilt modes. To correct other low-order aberrations as well and to improve the speed, we are currently upgrading the way the LLOWFS interacts with the PyWFS. The LLOWFS will send its corrections directly to the PyWFS that will then overwrite its reference point to compensate for these corrections with the DM.}
\item{The PyWFS, in its initial stage corrected only tip-tilt in the visible, which could not provide significant improvement for LLOWFS in the IR channel. Hence, the LLOWFS performance in both configurations was dominated by the uncorrected higher order modes.}
\end{itemize}
%
%
%
%
%  ** 5)  LLOWFS comaptibility **
%================================

\section{LLOWFS compatibility with coronagraphs}

The LLOWFS is compatible with a family of small IWA PMCs. 
Similar closed-loop laboratory performance has been obtained for the FQPM and the 8OPM coronagraphs as demonstrated for the VVC in the section~\ref{s:lab}. Detailed closed-loop performance analysis of the LLOWFS with different PMCs is intended for the future publication.

However, for coronagraphs such as the PIAA using an amplitude mask, the LLOWFS sensing capability depends on the size of the FPM. An amplitude mask bigger than the PSF core blocks most of the starlight and diffracts only a small fraction of it in the re-imaged pupil plane. LLOWFS, in that case, does not get enough starlight photons, hence cannot provide an optimal solution. However, we have closed the loop with PIAA and shaped pupil with an opaque binary FPM about half of the size of the PSF in the laboratory. For this type of coronagraphs, the CLOWFS would be a more efficient wavefront sensor. However, it requires some hardware changes on SCExAO that will not be compatible with the PMCs.

A way to make LLOWFS also efficient with the amplitude masks is to use a conic-shaped FPM that diffracts the starlight in a ring around the pupil in the Lyot plane. Such a mask provides an optimal  number of photons for LLOWFS independently of the size of the mask. We have tested this solution with an achromatic phase-shifting focal plane mask \citep[AFPM,][]{kevin}, which is based on a diffractive optical filtering technique scaling the size of the FPM linearly with the wavelength. This mask has a cone structure at its center with an angle optimized for the residual starlight to fall within the reflective zone of the RLS. The testing of AFPMs with PIAA and shaped pupil are currently ongoing on our instrument and the performance of the low-order correction in the laboratory and on the sky will be discussed in future publications.          
%
%
%
% ** 6) Conclusion **
%================================

\section{Conclusion}

Small IWA phase mask coronagraphs, which enables high contrast imaging at small angular separations, are extremely sensitive to tip-tilt errors. It is crucial to decrease these effects using all the rejected starlight available, which is typically discarded in a coronagraph. Hence, to overcome the consequences of wavefront aberrations at/near the diffraction limit, implementing LLOWFS-like technology is crucial to control starlight leakage around the coronagraphic mask. We have demonstrated the first successful on-sky closed-loop test of low-order corrections using LLOWFS with the vector vortex coronagraph on the SCExAO instrument.  

Both in the laboratory and on-sky, we showed an improvement of the low-order slow varying residuals ($<$~0.5~Hz), dynamically resolved by the exposure time of the science camera HiCIAO. In the laboratory, we obtained a correction of about 2 orders of magnitude for 35 Zernike modes, while on sky, due to the use of a small conservative gain for the controller, the improvement is only a factor of 3 for 10 Zernike modes. 

We also demonstrated the capacity of the low-order control loop to be combined with the higher-order loop for the correction of the non-common path and chromatic aberrations between this high-order loop and the coronagraph. We obtained a factor 3 to 4 improvement in a preliminary setup, using a slow differential pointing system. These results are expected to improve with the better integration of the low-order differential control in the high-order loop.

Corrections of higher-order modes other than just tip-tilt by PyWFS should provide a Strehl ratio $>~90\%$. Moreover, the implementation of a LQG control law in the low-order correction should significantly reduce the coronagraphic leakage in the IR channel. Further performance testing of the integrated control loop on-sky is scheduled for the upcoming observational nights at the Subaru Telescope. A significant enhancement in the detection sensitivity of the SCExAO instrument is expected during the future science observations. 
%
%
%
% ** 7) Future Development **
%================================

\section{Future Development}

Future work related to the LLOWFS is focused on three areas which are envisioned to provide high contrast at small angular separation. The goals are: optimal control of the low-order aberrations, point spread function calibration close to/near the IWA using low-order telemetry \citep{Vogt} and interaction between speckle calibration and low-order control. 

We are currently in the process of implementing a LQG controller for the LLOWFS on SCExAO. In order to improve the post processing of the science images, we will use the low-order telemetry of the residuals left uncorrected by the control loop to calibrate the amount of starlight leakage at small angular separations. We are also currently studying the interaction of speckle calibration with the LLOWFS especially for the correction of speckles at small IWA. 

The development and the implementation of the above mentioned technologies on SCExAO should significantly improve the contrast around the first couple of Airy disks of the star. Such advancements will allow SCExAO to detect young Jupiters (a few $M_{j}$) by a factor of $\sim$~3 closer to their host stars than is currently possible with other ground-based ExAO systems.

Our goal is to demonstrate innovative wavefront control approaches that are central to future high contrast systems. To maximize the performance of the coronagraphs by efficiently controlling and calibrating the wavefront at the small angular separations, we aim to search the best instrumental parameter space to combine the optimized LLOWFS control with the PSF calibration and speckle nulling. A precursor of these approaches implemented on the next generation extremely large telescopes and future larger space missions should enable direct imaging and low resolution spectroscopy of Earth-like planets in the HZ of M-type and F,G,K-type nearby stars respectively. 
%
%
%
% **8) Acknowledgment **
%================================

\section{Acknowledgment}

The SCExAO team would like to thank the AO188 scientists and engineers for operating the AO system and diagnosing the issues faced during the observations. We gratefully acknowledge the support and help from the Subaru Observatory staff. This research is partly supported by a Grant-in-Aid for Science Research in a Priority Area from MEXT, Japan.
%

%\section{Bibliography}


\begin{thebibliography}{30}
%\expandafter\ifx\csname natexlab\endcsname\relax\def\natexlab#1{#1}\fi

\bibliographystyle{mnras}
%\nocite{*}
%\bibliography{reference}

\bibitem[{Belikov} et~al.(2014){Belikov}, {Lozi}, {Pluzhnik}
  et~al.]{belikov}
{Belikov} R., {Lozi} J., {Pluzhnik} E., et~al., 2014, in { Society of
  Photo-Optical Instrumentation Engineers (SPIE) Conference Series\/}, vol.
  9143 of { Society of Photo-Optical Instrumentation Engineers (SPIE)
  Conference Series\/}, ~23

\bibitem[{Beuzit} et~al.(2010){Beuzit}, {Boccaletti}, {Feldt}
  et~al.]{sphere}
{Beuzit} J.-L., {Boccaletti} A., {Feldt} M., et~al., 2010, in { Pathways
  Towards Habitable Planets\/}, edited by V.~{Coud{\'e} du Foresto}, D.~M.
  {Gelino}, I.~{Ribas}, vol. 430 of { Astronomical Society of the Pacific
  Conference Series\/},  231

\bibitem[{Clergeon} et~al.(2013){Clergeon}, {Guyon}, {Martinache}
  et~al.]{chris}
{Clergeon} C., {Guyon} O., {Martinache} F., et~al., 2013, in { Proceedings of
  the Third AO4ELT Conference\/}, edited by S.~{Esposito}, L.~{Fini}, ~95
  
\bibitem[{Guyon}(2003)]{piaa}
{Guyon} O., 2003, \aap, 404, 379

\bibitem[{Guyon} et~al.(2006){Guyon}, {Pluzhnik}, {Kuchner}, {Collins} \&
  {Ridgway}]{wferror}
{Guyon} O., {Pluzhnik} E.~A., {Kuchner} M.~J., {Collins} B., {Ridgway} S.~T.,
  2006, \apjs, 167, 81
  
\bibitem[{Guyon} et~al.(2009{\natexlab{a}}){Guyon}, {Matsuo} \&
  {Angel}]{clowfs1}
{Guyon} O., {Matsuo} T., {Angel} R., 2009{\natexlab{a}}, \apj, 693, 75

\bibitem[{Guyon} \& {Martinache}(2013)]{elt}
{Guyon} O., {Martinache} F., 2013, in { American Astronomical Society Meeting
  Abstracts \#221\/}, vol. 221 of { American Astronomical Society Meeting
  Abstracts\/}
  
\bibitem[{Guyon} et~al.(2014){Guyon}, {Hayano}, {Tamura}
  et~al.]{ao}
{Guyon} O., {Hayano} Y., {Tamura} M., et~al., 2014, in { Society of
  Photo-Optical Instrumentation Engineers (SPIE) Conference Series\/}, vol.
  9148 of { Society of Photo-Optical Instrumentation Engineers (SPIE)
  Conference Series\/}, ~1

\bibitem[{Hodapp} et~al.(2008){Hodapp}, {Suzuki}, {Tamura}
  et~al.]{Hodapp}
{Hodapp} K.~W., {Suzuki} R., {Tamura} M., et~al., 2008, in { Society of
  Photo-Optical Instrumentation Engineers (SPIE) Conference Series\/}, vol.
  7014 of { Society of Photo-Optical Instrumentation Engineers (SPIE)
  Conference Series\/}
  
  \bibitem[{Jovanovic} et~al.(2015){Jovanovic}, {Guyon}, {Martinache}
  et~al.]{scexao}
{Jovanovic} N., {Guyon} O., {Martinache} F., et~al., 2015, submitted in PASP
  
  \bibitem[{Kasdin} et~al.(2004){Kasdin}, {Vanderbei}, {Littman}, {Carr} \&
  {Spergel}]{shaped}
{Kasdin} N.~J., {Vanderbei} R.~J., {Littman} M.~G., {Carr} M., {Spergel} D.~N.,
  2004, in { Optical, Infrared, and Millimeter Space Telescopes\/}, edited by
  J.~C. {Mather}, vol. 5487 of { Society of Photo-Optical Instrumentation
  Engineers (SPIE) Conference Series\/},  1312--1321

\bibitem[{Kern} et~al.(2013){Kern}, {Guyon}, {Kuhnert}, {Niessner},
  {Martinache} \& {Balasubramanian}]{kern}
{Kern} B., {Guyon} O., {Kuhnert} A., {Niessner} A., {Martinache} F.,
  {Balasubramanian} K., 2013, in { Society of Photo-Optical Instrumentation
  Engineers (SPIE) Conference Series\/}, vol. 8864 of { Society of
  Photo-Optical Instrumentation Engineers (SPIE) Conference Series\/}, ~0
  
  \bibitem[{Lozi} et~al.(2014){Lozi}, {Belikov}, {Thomas}
  et~al.]{JL}
{Lozi} J., {Belikov} R., {Thomas} S.~J., et~al., 2014, in { Society of
  Photo-Optical Instrumentation Engineers (SPIE) Conference Series\/}, vol.
  9143 of { Society of Photo-Optical Instrumentation Engineers (SPIE)
  Conference Series\/}, ~22
     
  \bibitem[{Macintosh} et~al.(2014){Macintosh}, {Graham}, {Ingraham}
  et~al.]{gpi}
{Macintosh} B., {Graham} J.~R., {Ingraham} P., et~al., 2014, Proceedings of the
  National Academy of Science, 111, 12661

\bibitem[{Mawet} et~al.(2009){Mawet}, {Serabyn}, {Liewer}
  et~al.]{vvc}
{Mawet} D., {Serabyn} E., {Liewer} K., et~al., 2009, Optics Express, 17, 1902

  \bibitem[{Mawet} et~al.(2010){Mawet}, {Serabyn}, {Liewer}, {Burruss}, {Hickey}
  \& {Shemo}]{mawet}
{Mawet} D., {Serabyn} E., {Liewer} K., {Burruss} R., {Hickey} J., {Shemo} D.,
  2010, \apj, 709, 53

  \bibitem[{Murakami} et~al.(2010){Murakami}, {Guyon}, {Martinache}
  et~al.]{murakami}
{Murakami} N., {Guyon} O., {Martinache} F., et~al., 2010, in { Society of
  Photo-Optical Instrumentation Engineers (SPIE) Conference Series\/}, vol.
  7735 of { Society of Photo-Optical Instrumentation Engineers (SPIE)
  Conference Series\/}, ~33
  
  \bibitem[{Newman} et~al.(2014){Newman}, {Belikov}, {Pluzhnik},
  {Balasubramanian} \& {Wilson}]{kevin}
{Newman} K., {Belikov} R., {Pluzhnik} E., {Balasubramanian} K., {Wilson} D.,
  2014, in { Society of Photo-Optical Instrumentation Engineers (SPIE)
  Conference Series\/}, vol. 9151 of { Society of Photo-Optical Instrumentation
  Engineers (SPIE) Conference Series\/}, ~5
  
\bibitem[{Poyneer} et~al.(2014){Poyneer}, {De Rosa}, {Macintosh}
  et~al.]{Poyneer}
{Poyneer} L.~A., {De Rosa} R.~J., {Macintosh} B., et~al., 2014, in { Society of
  Photo-Optical Instrumentation Engineers (SPIE) Conference Series\/}, vol.
  9148 of { Society of Photo-Optical Instrumentation Engineers (SPIE)
  Conference Series\/}, ~0
 
  \bibitem[{Petit} et~al.(2009){Petit}, {Conan}, {Kulcs{\'a}r} \&
  {Raynaud}]{lqg}
{Petit} C., {Conan} J.-M., {Kulcs{\'a}r} C., {Raynaud} H.-F., 2009, Journal of
  the Optical Society of America A, 26, 1307
\bibitem[{Rouan} et~al.(2000){Rouan}, {Riaud}, {Boccaletti}, {Cl{\'e}net} \&
  {Labeyrie}]{Rouan}
{Rouan} D., {Riaud} P., {Boccaletti} A., {Cl{\'e}net} Y., {Labeyrie} A., 2000,
  \pasp, 112, 1479
  
 \bibitem[{Petit} et~al.(2014){Petit}, {Sauvage}, {Fusco}
  et~al.]{Petit}
{Petit} C., {Sauvage} J.-F., {Fusco} T., et~al., 2014, in { Society of
  Photo-Optical Instrumentation Engineers (SPIE) Conference Series\/}, vol.
  9148 of { Society of Photo-Optical Instrumentation Engineers (SPIE)
  Conference Series\/}, ~0
  
\bibitem[{Serabyn} et~al.(2010){Serabyn}, {Mawet} \&
  {Burruss}]{eugene}
{Serabyn} E., {Mawet} D., {Burruss} R., 2010, \nat, 464, 1018

  \bibitem[{Singh} et~al.(2014{\natexlab{a}}){Singh}, {Martinache}, {Baudoz}
  et~al.]{Singh1}
{Singh} G., {Martinache} F., {Baudoz} P., et~al., 2014{\natexlab{a}}, \pasp,
  126, 586
  
\bibitem[{Singh} et~al.(2014{\natexlab{b}}){Singh}, {Guyon}, {Baudoz}
  et~al.]{Singh2}
{Singh} G., {Guyon} O., {Baudoz} P., et~al., 2014{\natexlab{b}}, in { Society
  of Photo-Optical Instrumentation Engineers (SPIE) Conference Series\/}, vol.
  9148 of { Society of Photo-Optical Instrumentation Engineers (SPIE)
  Conference Series\/}, ~48
  
\bibitem[{Vogt} et~al.(2011){Vogt}, {Martinache}, {Guyon}
  et~al.]{Vogt}
{Vogt} F.~P.~A., {Martinache} F., {Guyon} O., et~al., 2011, \pasp, 123, 1434
%

\end{thebibliography}
\end{document}